\documentclass[aps,pra,superscriptaddress,twocolumn,notitlepage,longbibliography]{revtex4-1}
\usepackage{soul}
\usepackage[usenames,dvipsnames]{color}
\usepackage[latin1]{inputenc}
\usepackage{amsmath}
\usepackage{amsfonts}
\usepackage{amssymb}
\usepackage{bbm}
\usepackage{graphicx}
\usepackage{epstopdf}
\graphicspath{ {images/} }
\usepackage{amsmath}
\usepackage[titletoc]{appendix}
\usepackage{mathrsfs}
\usepackage{braket}
\usepackage{hyperref}

\def \bn{\begin{align}}
\def \en{\end{align}}
\def \be{\begin{equation}}
\def \ee{\end{equation}}
\def \bea{\begin{eqnarray}}
\def \eea{\end{eqnarray}}
\def \ba{\begin{array}}
\def \ea{\end{array}}

\def \yd{^\dagger}
\def \nd{^{\vphantom{\dagger}}}

\newcommand{\average}[1]{\left\langle{#1}\right\rangle}
\newcommand{\adag}{a^{\dagger}}
\newcommand{\gdn}{\gamma_{\downarrow}}
\newcommand{\gphi}{\gamma_{\phi}}

\newcommand{\sz}{\average{\sigma^z}}
\renewcommand{\aa}{\average{a a}}
\newcommand{\ada}{\average{\adag a}}

\newcommand{\xx}{C^{xx}}
\newcommand{\yy}{C^{yy}}
\newcommand{\zz}{C^{zz}}
\newcommand{\xy}{C^{xy}}

\newcommand{\abx}{C^{\alpha\beta}}
\newcommand{\ax}{C^{ax}}
\newcommand{\ay}{C^{ay}}

\newcommand{\exax}{\average{a \sigma^x}}

\newcommand{\OrderExponent}{\beta}
\newcommand{\SuscExponent}{\gamma}
\newcommand{\FiniteSizeExponent}{\zeta}
\newcommand{\edt}[1]{{\color{black}{#1}}}
\newcommand{\pk}[1]{{\color{black}{#1}}}
\newcommand{\jk}[1]{{\color{black}{#1}}}

\DeclareMathOperator\Tr{\mathrm{Tr}}
\renewcommand{\vec}[1]{\boldsymbol{#1}}

\let\Re\relax \DeclareMathOperator\Re{\mathrm{Re}}%
\let\Im\relax \DeclareMathOperator\Im{\mathrm{Im}}%

\def \av#1{{\langle#1\rangle}}

\newcommand{\mycite}[1]{{\textsuperscript{\cite{#1}}}}

\begin{document}

\title{Introduction to the Dicke model: from equilibrium to nonequilibrium, and vice versa}

\author{Peter Kirton}
\affiliation{SUPA, School of Physics and Astronomy, University of St Andrews, St Andrews, KY16 9SS, United Kingdom}
\affiliation{Vienna Center for Quantum Science and Technology, Atominstitut, TU Wien, 1040 Vienna, Austria}

\author{Mor M. Roses}
\affiliation{Department of Physics and Center for Quantum Entanglement Science and Technology, Bar-Ilan University, Ramat Gan 5290002, Israel}

\author{Jonathan Keeling}
\affiliation{SUPA, School of Physics and Astronomy, University of St Andrews, St Andrews, KY16 9SS, United Kingdom}

\author{Emanuele G. Dalla Torre}
\affiliation{Department of Physics and Center for Quantum Entanglement Science and Technology, Bar-Ilan University, Ramat Gan 5290002, Israel}

\begin{abstract}
The Dicke model describes the coupling between a quantized cavity field and a large ensemble of two-level atoms. When the number of atoms tends to infinity, this model can undergo a transition to a superradiant phase, belonging to the mean-field Ising universality class. The superradiant transition was first predicted for atoms in thermal equilibrium \edt{and was recently realized with a quantum simulator made of} \jk{atoms in an optical cavity, subject to both dissipation and driving.} In this Progress Report, we offer an introduction to some theoretical concepts relevant to the Dicke model, reviewing the critical properties of the superradiant phase transition, and the distinction between equilibrium and nonequilibrium conditions.  In addition, we explain the fundamental difference between the superradiant phase transition and the more common lasing transition. Our report mostly focuses on the steady states of \jk{atoms in} single-mode optical cavities, but we also mention some aspects of real-time dynamics, as well as \edt{other quantum simulators}, including \jk{superconducting qubits, trapped ions, and  using spin-orbit coupling for cold atoms.  These reali\edt{z}ations differ in regard to whether they describe equilibrium or non-equilibrium system\edt{s}.}

%Progress reports to be submitted to Advanced Quantum Technologies. Initial deadline: April 30. All factors of $2$ are set by the Dalla-Diehl paper\mycite{dalladiehl}.
\end{abstract}

\maketitle

\makeatletter
\def\l@subsection#1#2{}
\def\l@subsubsection#1#2{}
\makeatother

\tableofcontents

\section{Historical background}
\label{sec:history}
Superradiance was first introduced in 1954 by Dicke to describe the emission of light by a large ensemble of atoms\mycite{dicke54}. Dicke considered $N$ two-level atoms that are initially prepared in their excited state. At a given time, one of the atoms decays by emitting a photon. This induces a chain reaction that leads to the decay of all the $N$ atoms and the emission of $N$ photons in free space. Dicke explained that if all the atoms are trapped within a fraction of a wavelength, the photons emitted will be indistinguishable. In this case, the emission processes will interfere constructively, giving rise to an electromagnetic field with amplitude proportional to $N$
and an energy density proportional to $N^2$. The scaling laws of this {\it transient} superradiance differ from the decay of $N$ independent atoms, where the light is emitted incoherently and has an energy density proportional to $N$.

In 1973, Hepp and Lieb\mycite{hepp1973equilibrium} discovered a different type of {\it steady-state} superradiance,  which occurs when the ensemble of atoms is coupled to the quantized mode of a cavity.  They considered the thermal equilibrium properties of the resulting Dicke model and demonstrated that it shows a continuous phase transition between a normal and a superradiant phase. To achieve a meaningful thermodynamic limit, Hepp and Lieb\mycite{hepp1973equilibrium} assumed that the coupling between the two level systems and the photon field decreases as $1/\sqrt{N}$. Under this assumption, in the normal phase, the number of photons $n$ does not grow with $N$, while in the superradiant phase, $n$ is proportional to $N$. The paper by Hepp and Lieb is written in a mathematical style, which was soon reformulated in a form more transparent to physicists by Wang and Hioe\mycite{wang73}.  Their analysis was later refined by Refs.~\cite{hioe1973phase,charmichael1973higher,duncan1974effect} who showed that the transition survives in the presence of counter-rotating terms, which however shift the position of the transition by a factor of 1/2.

\begin{figure}[t]
  \centering
  \includegraphics[width=3.2in]{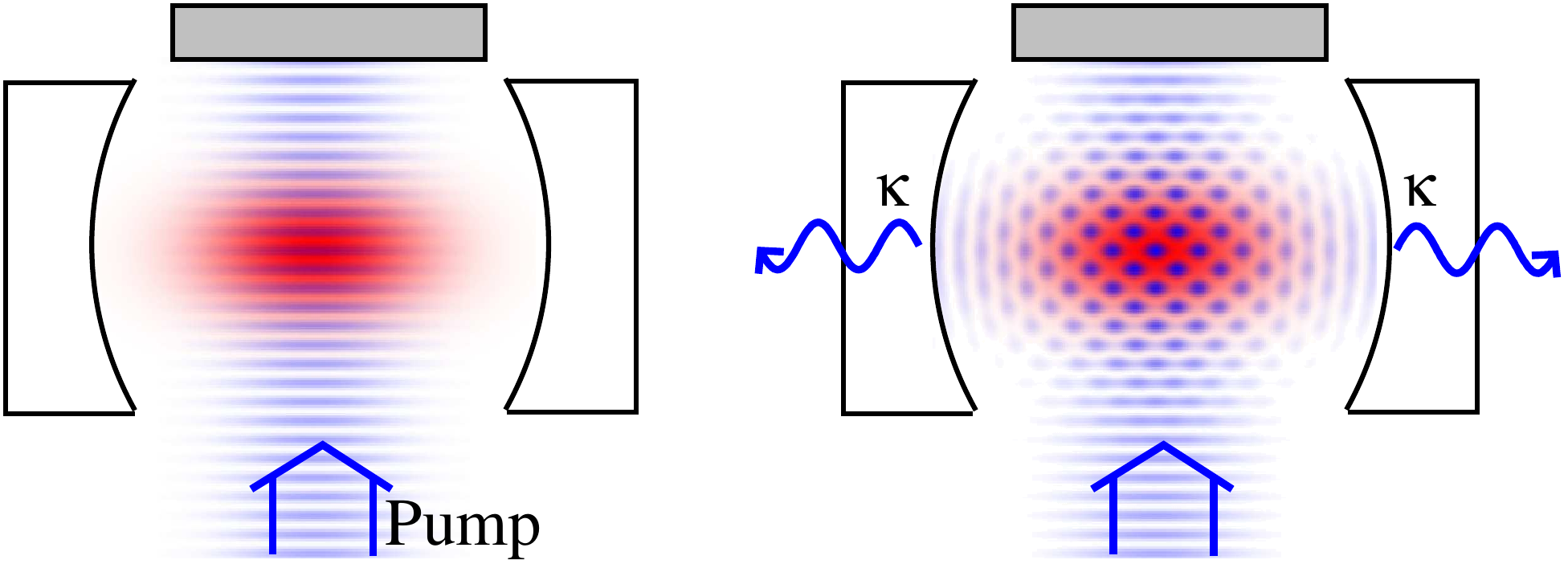}
  \caption{Cartoon of the self-organization transition. When the pump strength is below threshold (left), the atoms {are delocalized and scatter light incoherently in the cavity}. Above threshold (right) the atoms feel an optical lattice from the interference of pump and cavity light, and  organize into a checkerboard lattice. Adapted from Ref.~\cite{Keeling2010}.}
  \label{fig:self-org}
\end{figure}

In spite of the significant theoretical interest, the superradiant transition had not been realized experimentally, until recent times. {The major difficulty is that the transition requires very strong coupling between the atoms and the cavity, such that the photon-atom coupling is of the order of the atomic and cavity frequencies. From a theoretical \jk{perspective}, several authors studied whether the superradiant transition can be reached using only the dipole coupling between the atoms and the cavity. These studies gave rise to a fundamental debate around the validity of a no-go theorem for the superradiant transition, which will be discussed in Sec.~\ref{sec:other-real-dicke}.

\begin{figure}[t]
	\includegraphics[scale=0.5]{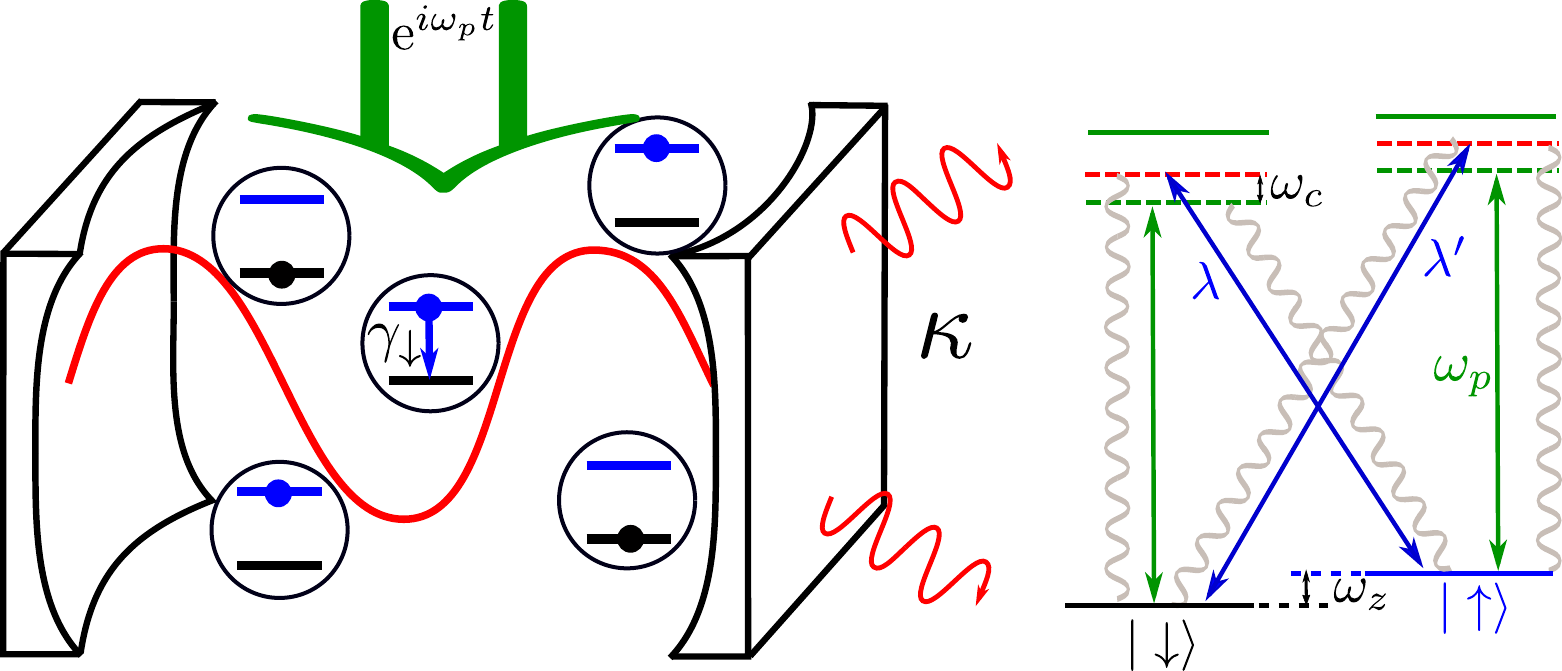}
	\caption{Schematic representation of the driven-dissipative Dicke model, based on internal degrees of freedom and proposed by Dimer \textit{et al.}\mycite{dimer07}. In this realization, each {atom} is modeled by a 4-level scheme and is coupled to the cavity through stimulated Raman emissions. In the steady state, the system absorbs energy from the external time dependent pump (at frequency $\omega_p$) and dumps it into several dissipative channels ($\gamma_{\downarrow}$ and $\kappa$).}
	\label{fig:schematic}
\end{figure}

In the last decade, two {uncontested} ways to realize the Dicke model and its superradiant transition have been demonstrated theoretically and experimentally. The first approach was proposed by Dimer \textit{et al.}\mycite{dimer07} and is based on a 4-level scheme ({see} Fig.~\ref{fig:schematic}). In this setup, the coupling between the atoms and the photons is induced by stimulated Raman emission, and can be made arbitrarily strong. This proposal was recently realized by Zhiqiang \textit{et al.}.\mycite{zhiqiang2017nonequilibrium}.
The second approach was inspired by an earlier experiment, proposed by Domokos and Ritsch\mycite{domokos2002collective}, and realized by Black \textit{et al.}\mycite{black03}. These authors considered a gas of thermal atoms that are trapped inside a cavity.  The atoms are illuminated by an external coherent pump and scatter photons into the cavity (see Fig.~\ref{fig:self-org}). It was found that for strong enough pump intensities, the atoms self-organize in a checkerboard pattern, where the atoms are preferentially separated by an integer multiple of the {photon's} wavelength, and {scatter} light coherently. This analysis was later extended to the case of a Bose-Einstein condensate (BEC) theoretically by Nagy \textit{et al.}\mycite{nagy10} and experimentally by Baumann \textit{et al.}\mycite{baumann10}. In a BEC, the atoms are delocalized, and the phase of the scattered light is random. In this situation, {the scattered photons are} incoherent and their number does not grow with $N$. In contrast, in the self-organized state, all atoms emit photons coherently, giving rise to a superradiant phase, where the number of photons is proportional to $N$. Following this reasoning, Refs.~\cite{nagy10,baumann10} showed that the onset of self-organization can be mapped to the superradiance transition of the Dicke model, see Sec.~\ref{sec:models}. This study was later extended to {narrow linewidth}\mycite{klinder2015dynamical} and multimode\mycite{vaidya18} cavities.}

The two above-mentioned realizations of the superradiant transition \jk{in the Dicke model} involve driven-dissipative systems. In both settings, the coupling between the atoms and the photons is achieved through an external time-dependent pump.  {This allows} {arbitrarily strong effective light-matter coupling strengths, enabling the transition}. As a consequence of being driven, these systems cannot be described by an equilibrium Dicke model, but one needs to take into account the drive and dissipation present. This subtle difference was initially dismissed because, in the limit of vanishing losses, the critical coupling of the driven-dissipative model coincides with the value of the equilibrium case, see Sec.~\ref{sec:critical}. Because the driven-dissipative model does not have a well-defined temperature, it was tempting to identify the experiment with a zero-temperature quantum phase transition. {However, later studies\mycite{nagy11,oeztop11,dalladiehl} showed that the phase transition has the same universal properties as {the} equilibrium transition at finite temperature. This equivalence can be understood in terms of an emergent low-frequency thermalization, which will be reviewed in Sec.~\ref{sec:universal}.} \edt{These} \jk{approaches can be considered as analog quantum simulators of the Dicke model:  the driving scheme is designed to engineer an effective Dicke model with tunable parameters allowing exploration of the phase diagram.  As discussed further in Sec.~\ref{sec:other-real-dicke}, there also exist proposals for digital or hybrid analog-digital quantum simulation of the Dicke model using superconducting qubits or trapped ions\mycite{mezzacapo2014digital,lamata2017digital,aedo2018analog}.}

%This identification was however questioned by Ref.~\cite{nagy11,oeztop11}, who found that the critical exponents of the driven-dissipative transition differ from the critical exponents of the quantum model.

The main goals of this Progress Report are (i) to present simple physical arguments to understand the commonalities and differences between the superradiant phase transition in the equilibrium Dicke model and its non-equilibrium counterparts (Secs.~\ref{sec:models}-\ref{sec:universal}), (ii) to introduce some analytical and numerical approximations, used to study the Dicke model (Sec.~\ref{sec:beyond}); and (iii) to set the superradiant transition in the wider context of closely related models and transitions \jk{(Sections~\ref{sec:lasing} and~\ref{sec:clos-relat-models})}. For a broader discussion of the phenomena of superradiance and the Dicke model, we refer the reader to a number of other relevant reviews:
Gross and Haroche\mycite{Gross1982a} discusses the transient superradiance first predicted by Dicke; Garraway\mycite{Garraway2011} presents the Dicke model and its phase transitions from a quantum optics perspective; Ritsch \textit{et al.}\mycite{Ritsch2013} discusses the
 self organization of atoms in optical cavities and dynamical optical lattices.

\section{Models and experiments}
\label{sec:models}

\subsection{The Dicke model at equilibrium}

The Dicke model describes a single bosonic mode (often a cavity photon mode) which interacts collectively with a set of $N$ two-level {systems} (the atoms). The Dicke Hamiltonian is given by
\begin{equation}
H = \omega_c a^\dagger a +\omega_z \sum_{j=1}^N \sigma^z_j +\frac{2\lambda}{\sqrt{N}} (a+a^\dagger) \sum_j \sigma^x_j\;. \label{eq:Dicke}
\end{equation}
Here $a^\dagger (a)$ are the creation (annihilation) operators of the photon, satisfying $[a,a^\dagger]=1$, and $\sigma^\alpha_i$ are spin operators, satisfying $[\sigma^x_j,\sigma^y_k]=i\delta_{j,k}\sigma^z_j$ (note that $\sigma^\alpha=\tau^\alpha/2$, where $\tau^\alpha$ are Pauli matrices). The model has three tuning parameters: the photon frequency $\omega_c$, the atomic energy splitting $\omega_z$, and the photon-atom coupling  $\lambda$.

To understand the nature of the superradiant transition, it is useful to analyze the symmetries of this model. By applying the transformation $a\to -a$ and $\sigma^x\to -\sigma^x$, the Hamiltonian remains unchanged. This gives a symmetry {group} with only two elements (when {this transformation is} applied twice it brings back to the original state) and is formally associated with a $\mathbb{Z}_2$ group. This symmetry arises due to the conservation of the parity of the total number of excitations (i.e. the number of photons, plus the number of excited spins), and is analogous to the Ising symmetry of ferromagnets. As we will see, the superradiant transition indeed shares the same critical exponents as the mean-field Ising transition.

The Dicke model, Eq.~(\ref{eq:Dicke}), depends on the atomic degrees of freedom through the total spin operators $S^\alpha=\sum_j \sigma^\alpha_j$ only. Using this definition, the Dicke model becomes
\be
\label{eq:H_totalS}
H = \omega_c a^\dagger a + \omega_z S^z + \frac{2\lambda}{\sqrt{N}} (a+a^\dagger)S^x\,.
\ee
This Hamiltonian commutes with the total spin $S^2 = (S^x)^2+(S^y)^2+(S^z)^2$. Consequently, it connects only states with the same total spin $S$, i.e. that belong to the same {\it Dicke manifold}. This symmetry provides a significant simplification of the problem because it allows the  description of the atomic degrees of freedom in terms of $N+1$ states, rather than the entire Hilbert space of size $2^N$\mycite{chen2008numerically}. This symmetry can however be broken by physical processes that act on individual atoms, which will be described in Sec.~\ref{sec:driven_models}.

\subsection{Raman transitions and self-organization}

As mentioned in the introduction, the Dicke model was realized experimentally in two ways: (i) using stimulated Raman emission between two hyperfine states in the ground state manifold of a cold atomic cloud, and (ii) coupling to the motional degrees of freedom of a BEC.

{The former realization\mycite{dimer07} involves a 4-level scheme, schematically drawn in Fig.~\ref{fig:schematic}.} The mapping to the Dicke model is straightforward: $\omega_z$ is the effective splitting between the two ground states (taking into account any differential Stark shifts due to the external drive), and $\lambda/\sqrt{N}$ the strength of the stimulated Raman emission into the cavity mode. {Note that this coupling is achieved by using two distinct external fields. These two processes correspond to $\sigma^+_i a + \sigma^-_i a^\dagger$ and $\sigma^+_i a^\dagger + \sigma^-_i a^\dagger$, respectively, and are often referred to as rotating and counter-rotating. When the two processes have equal strength, one recovers the Dicke model of Eq.~(\ref{eq:Dicke}). By varying the relative strength, it {is} possible to realize a generalized Dicke model, with different {prefactors to the} rotating and counter-rotating terms, which will be discussed \jk{further in} Sec.~\ref{sec:generalized}.}

In the latter realization\mycite{nagy10,baumann10}, the mapping to the Dicke model was achieved by considering two momentum modes of the atoms (the BEC at $q=0$ and the first recoil at $k_L=2\pi/\lambda$). {It is not immediately clear that this  mapping is completely justified}. Firstly, it is not \textit{a priori} clear that one may neglect higher order scatterings, at multiples of $k_L$. Secondly, {the mapping {only holds} if the atoms are} initially found in a BEC. However, in practice, the self-organization transition occurs in a thermal state as well\mycite{domokos2002collective,black03}: a detailed analysis revealed that the superradiance phase transition is essentially unaffected by the BEC transition\mycite{piazza2013bose}.

Hence, we present here a different mapping of the self-organization transition to the Dicke model, which does not require a BEC. Our derivation assumes that the atoms {do not interact} and are initially found in the superradiant phase. In this state, {the atoms scatter light into a standing wave of the cavity field, whose period is $\lambda/2$}. However, to enable superradiance, the atoms need to preferentially occupy sites {that are separated by an integer multiple of $\lambda$ in the longitudinal direction of the cavity. Having denoted all the possible sites as even or odd,  we introduce} the spin variables $\sigma^x_j$, which indicate whether the atom $j$ is on an even ($\sigma^x_j=1/2$) or odd ($\sigma^x_j=-1/2$) site.
%{If we assume that the atoms are distinguishable (for example due to a different longitudinal position),  the spin operators denote independent variables and commute among themselves}.
Depending on their positions, the atoms scatter light from the pump, and create cavity photons, with a phase of either $0$ or $\pi$. If we define
$N_{\rm even}$ and $N_{\rm odd}$ as the operators that count the number of atoms on the even and odd sites, respectively, the photon-atom coupling can be written as $ \lambda(t) a^\dagger(N_{\text{even}} - N_{\text{odd}}) + {\rm H.c.} = 2\lambda(t) a^\dagger \sum_j\sigma^x_j +{\rm H. c.}$, where $\lambda(t)=\lambda \exp(i\omega_p t)$ is proportional to the pump field and oscillates at the pump frequency $\omega_p$. In addition, the atoms can experience quantum tunneling between even and odd sites. This process is described by the spin-flip operator $\omega_z\sigma^z_j$, where $\omega_z$ is the tunnelling rate.

By combining these terms, we obtain the Dicke model
\begin{align}
H(t) = \omega_c a^\dagger a + \omega_z\sum_{j=1}^N \sigma^z_j +  2\left(\lambda(t) a + \lambda^*(t) a^\dagger\right) \sum_{j=1}^N\sigma^x_j.
\label{eq:Dicke_time}
\end{align}
In general, the parameters in this model may have a non-trivial dependence
on the pump strength. (For instance in a standing-wave pump profile, the tunneling matrix element is given by the difference of eigenvalues of the Mathieu equation. See the Appendix A.1 of Ref.~\cite{bhaseen12}.) On approaching the transition, the standing wave becomes weaker and $\omega_z$ achieves its maximal possible value, which equals to the recoil energy $E_R=k_L^2/2m$. In this limit, Eq.~(\ref{eq:Dicke_time}) becomes identical to the Dicke model obtained by Ref.~\cite{nagy10}, which started by considering a BEC of atoms.

\subsection{Driven-dissipative models}
\label{sec:driven_models}
The explicit time dependence of Eq.~(\ref{eq:Dicke_time}) can be removed by shifting to an appropriate rotating frame, i.e.\ by using the gauge transformation
%\begin{align}
$a \to e^{i\omega_p t} a$.
%\end{align}
This transformation brings Eq.~(\ref{eq:Dicke_time}) to the time-independent Dicke model, Eq.~(\ref{eq:Dicke}), with a renormalized cavity frequency $\omega_c\to\omega_c-\omega_p$. If the system were closed, this transformation would have no physical consequences. However, when the system is coupled to a bath, the transformation changes the properties of the bath, pushing it out of equilibrium. {In particular, {since all frequencies are renormalized down by $\omega_p$,} the transformation leads to a bath with both positive and negative frequencies, while equilibrium baths have positive eigenfrequencies only.}
%Because the optical frequency is the largest energy scale in the problem.
Hence, there are two equivalent ways to describe the driven-dissipative Dicke model: (i) in the laboratory frame, where the bath is in thermal equilibrium but the Hamiltonian is time dependent, and (ii) in the rotating frame, where the Hamiltonian is time independent, but the baths are effectively out of equilibrium.

In this report we follow the second, more common approach, and work in the rotating frame. Since the optical frequency is the largest scale in the problem, the baths can be approximated as Markovian\mycite{scully97}. As discussed for example in Ref.~\cite{dalladiehl}, Markovian baths generally violate the equilibrium fluctuation-dissipation relation. {This is because of the negative frequency bath components described above.}
%To see why this is the case, it is sufficient to recall that in the rotating frame, Markovian baths have both positive and negative eigenfrequencies.
These cannot be found at thermal equilibrium because their partition function is not normalizable (for a bath mode at frequency $\omega_b<0$, $Z=Tr[e^{\beta |\omega_b| a^\dagger a}]\to\infty$). In practice, this is not a problem because the occupation of the bath modes is actually set by their frequencies in the laboratory frame $\omega_p+\omega_b>0$, rather than in the rotating frame, $\omega_b<0$.

For optical frequencies at room temperature, the occupation of the bath modes can be safely approximated to zero, giving rise to the Lindblad-form master equation
\begin{equation}
  \label{eq:Lindblad}
	\dot\rho = -i[H,\rho]+\sum_i\gamma_i\mathcal{D}\left[L_i\right]
\end{equation}
where $\rho$ is the system's density matrix, and
\be \mathcal{D}\left[L\right] \equiv 2 L  \rho L^\dagger-\left\{L^\dagger L,  \rho\right\}.\label{eq:Lindblad2}\ee
Physically, the rates $\gamma_i$ and operators $L_i$ correspond to different sources of dissipation.
For experiments on the Dicke model, the most relevant sources of dissipation are listed in Table \ref{table:dissipation}, and can be divided  in two main categories: collective effects ($\kappa$ and $\gamma$) and single atoms effects ($\gamma_\downarrow$ and $\gamma_\phi$). In Sec.~\ref{sec:critical} we will explain how to deal with these categories. Other sources of dissipation, such as the loss of atoms, require going beyond the picture of a fixed number of two-level systems coupled to light, and will not be considered here.

\begin{table}
\begin{tabular}{|c| c| c|}
\hline
{\rm rate} & L operator & {\rm physical process}\\
\hline
$\kappa$ & $a$ & cavity decay\\
\hline
$\gamma$ & $\sum_j \sigma^-_j=S^-$ & collective atomic decay\\
\hline
$\gamma_\downarrow$ & $\sigma^-_j$ & single-atom decay\\
\hline
$\gamma_\phi$ & $\sigma^z_j$ & single-atom dephasing\\
\hline
\end{tabular}
\caption{Main sources of dissipation that were considered in the literature\mycite{dalla2016dicke,kirton2017suppressing,zhiqiang2018dicke} \label{table:dissipation}.}
\end{table}

\subsection{Other realizations of the Dicke model}
\label{sec:other-real-dicke}

{In Sec.~\ref{sec:history}, we mentioned a no-go theorem for the superradiant transition by Rzazewski\mycite{rzazewski1975phase}. These authors claimed that the superradiant transition cannot be reached using dipole couplings between atoms and photons. The key observation of Rzazewski\mycite{rzazewski1975phase} is that the Dicke model is incomplete, because it is not invariant under gauge transformations of the electromagnetic field. {A minimal change which recovers this invariance is to add a term} proportional to the square of the vector potential. The Thomas-Reiche-Kuhn sum {rule then} implies that the strength of this additional term is exactly that needed to inhibit the phase transition, leading to a ``no-go'' theorem\mycite{bialynicki1979no,keeling2007coulomb}.

The validity of this no-go theorem is still debated. In particular, a {full quantum treatment of the problem} requires not only the $A^2$ terms, but a description of the longitudinal Coulomb interactions between dipoles.  By considering a full description of a realistic system of atoms in a real cavity, Refs.~\cite{vukics2012adequacy,Vukics2014,vukics2015fundamental,griesser16,de2018cavity} showed that a phase transition can occur in the right geometry.
{Since the ``photon creation'' operator describes different physical fields in different gauges, it is important to check what physical fields acquire macroscopic expectations in such a transition.  Such analysis reveals that}
this transition is adiabatically connected to a crystalline transition  for motional degrees of freedom\mycite{vukics2015fundamental}, or to a ferroelectric transition for dipole couplings\mycite{de2018cavity}.
{Very recent works\mycite{Bernadis2018breakdown,Stokes2018gauge} have also noted that since the two-level approximation has \pk{a} different meaning in different gauges, its validity at strong coupling is not gauge invariant:  as such\mycite{Bernadis2018breakdown} shows that only in the dipole gauge can the two-level approximation be trusted.}
The question of how to properly describe matter--light coupling has also recently been discussed in the context of combining cavity quantum electrodynamics with density functional theory\mycite{flick17,rokaj18}.}

\jk{The reali\edt{z}ation of the Dicke model using Raman driving circumvents the no-go theorem\edt{, for the following} reasons\edt{:}  Firstly, the effective matter-light couplings appearing in this Hamiltonian are a combination of the bare coupling, the pump strength and the detuning.  As such, these are not subject to any oscillator\edt{-}strength sum\edt{-}rule.  Moreover, even the bare couplings appearing in the effective coupling relate to transitions between ground and \pk{excited atomic} states, rather than direct transitions between the low energy states forming the two-level system.  \edt{Finally,} the effective cavity frequency is tunable through the pump-cavity detuning.  As a result of all of these points, there is no longer any constraint \pk{on} the relation between \edt{the} parameters \edt{of} the model, and a superradiant transition is possible.  An $A^2$ term may nonentheless be present, but \pk{\edt{the system's} parameters can be chosen such that \edt{this term} is weak enough} to be ignored.}

In addition, the original equilibrium superradiant transition of the Dicke model is possible in a grand canonical ensemble\mycite{eastham00,eastham01}. In such an ensemble, one minimizes the grand potential $\Phi=- k_B T \ln(Z_{\rm GC})$, where $Z_{\rm GC} = {\rm Tr} \left[\exp(-\beta(H-\mu N_{\rm ex}))\right]$, and $N_{\text{ex}} = a^\dagger a + \sum_j \sigma^z_j + 1/2$.   The chemical potential $\mu$ shifts the effective parameters $\omega_c, \omega_z \to \omega_c-\mu, \omega_z-\mu$ such that the sum rule required for the no-go theorem no longer holds. Considering this ensemble only makes sense if the Hamiltonian preserves the number of excitations, i.e. working in the limit where counter-rotating terms can be dropped, giving rise to the Tavis--Cummings model (see Sec.~\ref{sec:TC}). Conceptually, this corresponds to considering a perfect cavity prepared with an initial finite excitation density and then asking for the ground state. This model can also describe the Bose--Einstein condensation of exciton-polaritons --- superpositions of microcavity photons and excitons\mycite{Deng2010a,Carusotto2013a} --- in the limit of a very good cavity\mycite{Kasprzak2006,Sun2017}.

Another context in which the Dicke transition is expected to be possible involves circuit QED\mycite{houck2012chip}.  Here, the two-level atoms are replaced by superconducting qubits, coupled to a common microwave resonator. \jk{This again can be considered as an analog quantum simulator, with the superconducting qubits acting as tunable artificial atoms.}  There has been much discussion on whether {the Hamiltonian describing such a system should \jk{contain} $A^2$ terms, and as such, whether it} is subject to the no-go theorem\mycite{Nataf2010,Viehmann2011,ciuti12,lambert2016superradiance,Bamba2016,Jaako2016,bamba17}.  For at least some designs of circuit, if one starts from the classical
Kirchoff equations (i.e. conditions on the currents and voltages) of the circuit, and proceeds to quantize these equations, the resulting Hamiltonian \jk{is not necessarily} subject to the no-go theorem.  i.e., there are cases where either the $A^2$ term is absent, or where it is present, but with a weaker coupling strength than required to prevent the phase transition.

The above realizations of the Dicke model involve coupling to a photonic mode, at optical or microwave frequencies. In addition, the Dicke model can be realized in any case where many spin degrees of freedom couple to a common bosonic mode.  There have been several proposals for realizing such a model where the bosonic mode corresponds to motion in an harmonic trap, i.e.\ a mechanical phonon mode, rather than a photon.

\jk{
One widely studied example involves coupling the electronic states of trapped ions to their center of mass motion\mycite{genway2014generalized,pedernales2015quantum,aedo2018analog,safavi2018verification}.  In fact, the natural coupling between a standing wave laser and an ion leads to a position dependent matrix element\mycite{leibfried03}.  Writing this position in terms of vibrational raising and lowering operators, one can expand in the Lamb-Dicke regime to produce an effective Dicke model\mycite{pedernales2015quantum,aedo2018analog}.
Alternately, a state-dependent optical potential can be used to couple the electronic state of the ion to the center of mass mode\mycite{porras04,wang13,genway2014generalized}.  Such an approach has been realized experimentally in Ref.~\cite{safavi2018verification}, where an adiabatic sweep from the normal to the superradiant state has been studied.

A similar idea has also been realized by Hamner \textit{et al.}\mycite{hamner2014dicke}, using a spin-orbit coupled BEC in an harmonic trap. Here spin-orbit coupling produces a coupling between atomic motion and the internal spin state.  The cloud of atoms is reduced to a single motional degree of freedom by the non-fragmentation of an interacting BEC.  Using this mapping to the Dicke model, the experimentally observed transition between a polarized and unpolarized state of the atoms can be understood as \pk{an} analogue of the superradiant phase transition.

All the above examples describe various routes to analog quantum simulation of the Dicke model, i.e.\ they involve directly engineering a Dicke Hamiltonian, and then studying the steady state or dynamics of this model. In addition, there have been other proposals to use digital quantum simulation, i.e.\ to
replace time evolution under the Dicke Hamiltonian with a sequence of
discrete unitary gates that leads to the same evolution.  In particular,
schemes have been proposed to reali\edt{z}e such digital quantum simulation using
superconducting qubits\mycite{mezzacapo2014digital,lamata2017digital}.
}

\section{Threshold of the superradiant transition}
\label{sec:critical}

In this section we give an overview of some simple techniques for finding the critical point in the Dicke model both in and out of equilibrium. These approaches are based on  mean-field theory, and give an intuitive understanding of the superradiant transition.

\subsection{Equilibrium transition}
\label{sec:MF}
In equilibrium we can calculate the critical coupling of the Dicke model, Eq.~(\ref{eq:Dicke}), by minimizing its mean-field free energy. Within this approach, we assume the photons to be in a coherent state $|\alpha\rangle$, defined by $a |\alpha\rangle = \alpha |\alpha\rangle$, where $\alpha$ is a real variational parameter. In this state, the energy of the cavity is $\omega_c\av{a\yd a}=\omega_c\alpha^2$ and each atom experiences the Hamiltonian
\be h(\alpha)=\omega_z \sigma_i^z + \frac{4\lambda}{\sqrt{N}}\alpha\sigma_i^x\;.\ee
The partition function is then given by
\begin{equation}
	Z(\alpha) = {\rm Tr}[e^{-\beta H}]={\rm e}^{-\beta\omega_c\alpha^2}\left(\Tr {\rm e}^{-\beta h}\right)^N\;,
\end{equation}
where $\beta=1/T$ is the inverse temperature. By definition, the free energy is \be F (\alpha)= -\frac1\beta \ln(Z(\alpha)) = \omega_c\alpha^2 - \frac{N}{\beta} \ln\left(2\cosh {\beta E}\right)\,,\label{eq:mft-f}\ee
 where
$E=\sqrt{\frac{\omega_z^2}{4} + \frac{4\lambda^2}N \alpha^2}$ is the eigenvalue of $h(\alpha)$.

By optimizing $F$ as a function of $\alpha$, one finds that if $\lambda<\lambda_c$ the minimum is at $\alpha=0$ while for $\lambda>\lambda_c$ the minimum is at $\alpha\neq 0$. The critical value $\lambda_c$ is found by the condition $F''(\alpha=0)=0$, or
\begin{equation}
 \lambda_c = \frac12\sqrt{\omega_c\omega_z\coth\left(\frac{\beta\omega_z}{2}\right)}.
 \label{eq:gc_MF}
\end{equation}
Note that this critical coupling smoothly evolves down to zero temperature  ($\beta\to\infty$), where one obtains $\lambda_c = \sqrt{\omega_c\omega_z}/2$. 

\jk{One may also use the above approach to find the critical exponent $\OrderExponent$ that controls how the order parameter $\alpha$ evolves beyond the critical $\lambda$.  In general, for $\lambda>\lambda_c$ we minimi\edt{z}e the free energy by solving \edt{$dF(\alpha)/d\alpha=0$, or}
\begin{equation}\label{eq:sc-condition}
  \omega_c  \alpha = \frac{N}{2} \tanh(\beta E) \frac{d E}{d \alpha}
  =  \frac{2\lambda^2 \tanh(\beta E)}{E}  \alpha,
\end{equation}
and since $\alpha \neq 0$ this gives $\omega_c E = 2 \lambda^2 \tanh(\beta E)$.  For small $\alpha$ we can expand $E = (\omega_z/2) + 4 \lambda^2 \alpha^2 / N \omega_z$.  Expanding both sides of Eq.~(\ref{eq:sc-condition}) to order $\alpha^2$, one  finds
\begin{equation}
  \label{eq:beta-exponent-def}
  \alpha = \sqrt{N \mathcal{A}(\lambda) (\lambda^2 - \lambda_c^2)},
\end{equation}
where $\mathcal{A}(\lambda)$ is a function of coefficients which is finite at $\lambda=\lambda_c$ for all temperatures.  \edt{One finds that} in the superradiant state, the order parameter scales as $\sqrt{N}$ \edt{and develops as $\alpha \sim (\lambda-\lambda_c)^\OrderExponent$, with} $\OrderExponent=1/2$. \edt{These results are valid both} for zero and non-zero temperature\edt{s}.}
Nevertheless, as we will explain in Sec.~\ref{sec:universal}, the \edt{these two} transitions are actually fundamentally different.

\subsection{Holstein--Primakoff transformation}
\label{sec:HP} An alternative description of the Dicke model relies on the Holstein-Primakoff \jk{(HP) representation}\mycite{holstein40}, which maps the total spin operators $S^\alpha$ to a bosonic mode $b$
\begin{align} \label{eq:HP}
S^z\to -\frac{N}2 + b^\dagger b ,~~~~~S^+
\to b^\dagger \sqrt{{N}- b^\dagger b}\,.
\end{align}
In the large $N$ limit (where $N\gg\av{b^\dagger b}$),  Eq.~(\ref{eq:HP}) simplifies to $S^x \to \sqrt{N}(b+b^\dagger)$ and the Dicke model, Eq.~(\ref{eq:H_totalS}), becomes equivalent to two coupled Harmonic oscillators
\be H_{HP} = \omega_c a^\dagger a + \omega_z b^\dagger b + \lambda (a+a^\dagger)(b+b^\dagger) \;.
\label{eq:H_MF}
\ee
Since the HP transformation relies on the total spin representation, this approach can include collective decay channels only, $\kappa$ and $\gamma$ in Table~\ref{table:dissipation}\mycite{Gelhausen2017}. Being a quadratic Hamiltonian, the model (\ref{eq:H_MF}) can be analytically solved at equilibrium, as well as out of equilibrium, in many different ways. In the following sections we will briefly summarize how this is done using \pk{m}aster equations, as well as Keldysh path integrals.

Within the \pk{m}aster equation approach, Eq.~(\ref{eq:Lindblad}), one has
\begin{equation}
\label{eq:master}
	\dot\rho=-i[H_{HP}, \rho] + \kappa\mathcal{D}[a] + \gamma\mathcal{D}[b].
\end{equation}
Eq.~(\ref{eq:master}) gives rise to linear equations of motion for the operators $a$ and $b$, which can be equivalently rewritten in terms of classical expectations,
\begin{align}
        \label{eq:da}
  \dot{a}&= (-i\omega_c - \kappa) a - i \lambda(b+b^\dagger) \\
        \label{eq:db}
  \dot{b}&= (-i\omega_z - \gamma) b - i \lambda(a+a^\dagger).
\end{align}
{These equations can be written in a matrix notation as
\begin{align}
        \label{eq:lineom}
        \dot{\vec{v}}(t)=&M\vec{v}(t),
\end{align}
with $\vec{v}=(a,a\yd,b,b\yd)^T$ and
\begin{align}
        M=&\begin{pmatrix}
                -(\kappa+i\omega_c)&0&-i\lambda&-i\lambda\\
                0&-(\kappa-i\omega_c)&i\lambda&i\lambda\\
                -i\lambda&-i\lambda&-(\gamma+i\omega_z)&0\\
                i\lambda&i\lambda&0&-(\gamma-i\omega_z)
        \end{pmatrix}.
        \label{eq:M_HP}
\end{align}

\jk{We now relate this expression to the retarded Green's function and the Keldysh path integral formalism.  The Keldysh formalism allows one to extend path integrals to systems away from thermal equilibrium.  Many comprehensive introductions to this approach can be found in textbooks\mycite{altlandbook,kamenev_book} as well as reviews of its application to driven-dissipative systems\mycite{sieberer2016keldysh}.  Given these excellent introductions, we do not aim here to discuss the derivation of this path integral, but provide a brief summary of its significance instead.} \edt{The key feature of the Keldysh formalism is the separate treatment of the retarded/advance\pk{d} Green's function, $G^{R/A}$, and the Keldysh Green's function, $G^K$. The former describe the response of the system to an external drive, while the latter captures thermal and quantum fluctuations inherent to the system. At thermal equilibrium these two quantities are linked by the the fluctuation-dissipation relations, which become invalid in the presence of external time-dependent drives. Formally, the distinction between $G^{R/A}$ and $G^K$ is achieved by the introduction of two separate fields that describe the evolution of the left (ket) and right (bra) side of the density matrix, respectively.}

As explained in Appendix A, Eq.~\eqref{eq:lineom} can be used to derive the retarded Green's function of the system
\begin{align}
        \label{eq:irgfo}
        \left[G^R(\omega)\right]^{-1}=&S^{-1}\left(\omega-iM\right),
\end{align}
here $S$ represents the equal-time commutation relations $S_{i,j}=\average{\left[v_i\nd(0),v_j\yd(0)\right]}$ and in the present case is given by:
\begin{align}
        \label{eq:Sv}
        S=\text{diag}(1,-1,1,-1).
\end{align}
Plugging Eqs.~\eqref{eq:M_HP} and \eqref{eq:Sv} into Eq.~\eqref{eq:irgfo} one finds
{%Old version
}%End of old version
%\begin{widetext}
\begin{multline}
{\left[G^R_{HP}(\omega)\right]}^{-1} =\\% M - \omega S =\\
\begin{pmatrix}
  \omega - \omega_c + i \kappa & 0  & -\lambda & -\lambda \\
0 & - \omega - \omega_c - i \kappa & -\lambda & -\lambda \\
-\lambda & -\lambda &\omega - \omega_z +i\gamma  & 0  \\
-\lambda & -\lambda & 0&- \omega - \omega_z -i\gamma  \\
\end{pmatrix}\label{eq:GR_HP}
\end{multline}}
In the limit of $\gamma\to0$, this expression is equivalent to the retarded Green's function derived in Ref.~\cite{dalladiehl}. %(Eq.(48))
%\end{widetext}

The superradiant transition corresponds to the requirement that one of the eigenvalues of $M$ goes to zero, or equivalently that  ${\rm det}[G^{R}(\omega=0)]=0$. This condition can be easily evaluated to deliver
\be
\lambda_c = \frac12\sqrt{\frac{\omega_z^2+\gamma^2}{\omega_z}\frac{\omega_c^2+\kappa^2}{\omega_c}}
\;.
\label{eq:gc_collective}
\ee
In the limit $\kappa,\gamma \to 0$, Eq.~(\ref{eq:gc_collective}) recovers the zero temperature limit of the equilibrium result, Eq.~\eqref{eq:gc_MF}. However, as we will explain in Sec.~\ref{sec:universal}, the transition of the open system is in a different universality class than the zero temperature limit.

\subsection{Critical coupling in the presence of single-atom losses}
The Holstein-Primakoff approximation assumes that the total spin of the model is conserved. As a consequence, it cannot describe processes that act on individual atoms, such as the single-atom decay $\gamma_{\downarrow}$ and dephasing $\gamma_\phi$ mentioned in Sec.~\ref{sec:driven_models}.
The effect of these processes on the critical coupling can be found by considering the equations of motion for the expectation values of the physical observables. Starting from the Hamiltonian in Eq.~(\ref{eq:Dicke}) and including the single atom decay sources, one finds\mycite{kirton2017suppressing}:
\begin{align}
  \label{eq:mfa}
  \partial_t \average{a} &= -\left(i\omega_c+\kappa\right)\average{a} - i{2\lambda\sqrt{N}}\langle \sigma^x\rangle \\
  \label{eq:mfb}
  \partial_t \langle \sigma^+\rangle &= (i\omega_z-\gamma_T) \langle \sigma^+\rangle  -\tfrac{2i\lambda}{\sqrt{N}}\Re[\langle a \sigma^z\rangle]
\end{align}
where $\gamma_T=\gphi+\gdn$. The above equations are exact, but do not form a closed set due to the terms $\langle a \sigma^z \rangle$. However, in the mean-field limit one can assume this factorizes as
$\langle a \sigma^z \rangle = \langle a \rangle \langle \sigma^z \rangle$.
This produces a closed set of mean field equations which are analogous to
the Maxwell-Bloch (MB) classical theory of a laser\mycite{Haken1970}.

%{\tt JK Added mention of Maxwell-Bloch (MB) equations}

The critical coupling of the superradiant transition can be found through a  linear stability analysis of Eqs.~(\ref{eq:mfa}) and (\ref{eq:mfb})\mycite{kirton2017suppressing}: By retaining only terms that are linear in $\langle a \rangle$ and $\langle \sigma^+ \rangle$,  {one obtains the same form as Eq.~(\ref{eq:lineom}), with}
% $i dv/dt=M_{MB} v$, with
\begin{multline}
  M_{\rm MB} =\\  {\begin{pmatrix}
  -(\kappa+i\omega_c)&0&-i\lambda&-i\lambda\\
  0&-(\kappa-i\omega_c)&i\lambda&i\lambda\\
  2i\lambda \av{\sigma^z}&2i\lambda\av{\sigma^z}&-(\gamma_T+i\omega_z)&0\\
  -2i\lambda\av{\sigma^z}&-2i\lambda\av{\sigma^z}&0&-(\gamma_T-i\omega_z)
  % -\omega_c + i \kappa & 0  & -\lambda & -\lambda \\
  % 0 &-\omega_c - i \kappa & -\lambda & -\lambda \\
  % -2\lambda\av{\sigma^z} & -2\lambda\av{\sigma^z} & -\omega_z +i\gamma_T  & 0  \\
  % -2\lambda\av{\sigma^z} & -2\lambda\av{\sigma^z} & 0& -\omega_z -i\gamma_T \\
\end{pmatrix}\;,}
\label{eq:M_MB}
\end{multline}
{and $\vec{v}=(\av{a}, \av{a^\dagger}, \av{\sigma^-}, \av{\sigma^+})^T$.}
% \begin{align}
% M_{\rm MB} &=  \left(\begin{array}{cccc}
% \omega_c - i \kappa & 0  & \lambda & \lambda \\
% 0 &- \omega_c - i \kappa & -\lambda & -\lambda \\
% 2\lambda\av{\sigma^z} & 2\lambda\av{\sigma^z} & \omega_z -i\Gamma  & 0  \\
% -2\lambda\av{\sigma^z} & -2\lambda\av{\sigma^z} & 0& -\omega_z -i\Gamma  \\
% \end{array}\right)\;,\label{eq:M}
% \end{align}
%\texttt{Check change $\gamma$ to $\Gamma$ above}
%\end{widetext}
%$v=\big(\av{a},~\av{a^\dagger},~\av{\sigma^-}/\sqrt{N},~\av{\sigma^+}/\sqrt{N}\big)^T$, $\sigma^\pm=\sigma^x\pm i \sigma^y$, and $\av{\sigma^z}=\av{S^z}/N$.
The superradiant transition occurs when the determinant of the above matrix vanishes, or equivalently:
\be \lambda_c =\frac12 \sqrt{\frac{(\omega_z^2+\gamma_T^2)(\omega^2_c+\kappa^2)}{-2\av{\sigma^z}\omega_z\omega_c}}.\label{eq:gc_phi}\ee
Note that if the atoms are initially fully polarized in the down state, i.e. $\av{\sigma^z}=-1/2$, then Eqs.~(\ref{eq:M_MB}) and (\ref{eq:gc_phi}) become equivalent to Eqs.~(\ref{eq:M_HP}) and (\ref{eq:gc_collective}).

\section{Universality in and out of equilibrium}
\label{sec:universal}

In this section we describe the critical properties of the superradiant transition, {from a theoretical {perspective}}: We first review the results obtained for the Dicke model at equilibrium (\ref{sec:critical_eq}) and out-of-equilibrium (\ref{sec:critical_neq}), and then explain the universal nature of these results in terms of analogous models of simple nonlinear oscillators (\ref{sec:Landau}).

\subsection{Equilibrium transition of the Dicke model}
\label{sec:critical_eq}

For a closed system at zero temperature, physical quantities in the normal phase of the Dicke model can be computed directly from the quadratic model of Eq.~(\ref{eq:H_MF}). This Hamiltonian can be diagonalized using a Bogoliubov transformation. For simplicity, let us consider the specific case of $\omega_c=\omega_z=1$. In this case, the Hamiltonian (\ref{eq:H_MF}) can be written as
\be
H = \frac12(p_a^2 + p_b^2) + \frac12(x_a ~x_b)\left(\begin{array}{c c}1 & 2\lambda \\ 2\lambda & 1\end{array}\right)\left(\begin{array}{c}x_a \\ x_b\end{array}\right)
\ee
where $x_a=(a+a\yd)/\sqrt{2}$ and $p_a=i(a\yd-a)/\sqrt{2}$. This Hamiltonian is diagonalized by the eigenmodes $x_\pm = (x_a \pm x_b)/\sqrt{2}$ and $p_\pm = (p_a \pm p_b)/\sqrt{2}$, with eigenfrequencies $\omega_\pm =1\pm 2\lambda$. In the new basis, the Hamiltonian decouples into two independent harmonic oscillators: $H_\pm=(p_\pm^2+\omega_\pm^2 x_\pm^2)/2$. The superradiant transition occurs when one of $\omega_\pm=0$, or equivalently $|\lambda|=\lambda_c=1/2$, as predicted by Eq.~(\ref{eq:gc_MF}).

{Let us now consider separately {the zero and finite temperature cases}. In the former case, one needs to calculate the ground state of an harmonic oscillator, where} $\av{x_\pm^2}=1/\sqrt{2\omega_\pm}$, leading to
\be \av{x_a^2} = \av{x_+^2}+\av{x_-^2} =  \frac1{2\sqrt{\lambda_c+\lambda}}+\frac1{2\sqrt{\lambda_c-\lambda}}.\label{eq:aa_eq}\ee
We can use this result to compute the critical exponent $\SuscExponent$, defined by $\av{a\yd a}\sim |\lambda-\lambda_c|^{-\SuscExponent}$. {The number of photons is $\av{a\yd a}=(\av{x_a^2}+\av{p_a^2}-1)/2$, {where $\av{x_a^2}$ diverges at the transition according to Eq.~(\ref{eq:aa_eq}), while $\av{p_a^2}$ remains finite. Consequently, the number of photons diverges as $(\lambda_c-\lambda)^{-1/2}$, leading to $\SuscExponent=1/2$.}

For a system at a finite temperature $T$, one has $\av{x_\pm^2}=
\coth(\beta  \sqrt{(\lambda_c\pm\lambda)/2})/(2\sqrt{(\lambda_c\pm\lambda))}$.  When the temperature is high compared to the mode frequency (which is always the case near the transition for the mode with vanishing frequency), one can approximate
$\av{x_\pm^2}=T/(\sqrt{2}(\lambda\pm\lambda_c))$, leading to the critical exponent $\SuscExponent=1$. These critical exponents are valid for any value of the $\omega_c/\omega_z$ ratio and demonstrate the difference between mean-field phase transitions at zero and finite temperatures.

\subsection{Non-equilibrium transition of the Dicke model}
\label{sec:critical_neq}

For a driven-dissipative model, it is necessary to use non-equilibrium techniques. Within the HP approximation, one obtains a quadratic Keldysh action of the form\mycite{dalladiehl}
\begin{align}\label{eq:Snormal}
S_{\rm N} =\frac{1}{2}\int_{\omega} V^\dag
\left(\begin{array}{cc}
0 & {[G^A_{HP}]}^{-1} \\
 {[G^R_{HP}]}^{-1} &  D_{HP}^K
\end{array}\right)V\;.
\end{align}
Here $V=(\vec{v};\bar{\vec{v}})$, where $\vec{v}$ is defined above and $\bar{\vec{v}}$ are auxiliary fields that allow us to describe the occupation of the bosons.
%At thermal equilibrium $D^K$ is determined by the fluctuation-dissipation theorem
%\be
%D^K(\omega) = 2{\rm Im}[G^R(\omega)] {\rm coth}\left(\frac{\omega}{2T}\right)
%\ee
For Markovian baths, $D^K$ is frequency independent and, if considering just photon loss, one simply has:
\be
D_{HP}^{K}= 2 \mathrm i \,\,\mathrm{diag} (\kappa,\kappa,0,0).\label{eq:G4x4}
\ee
By inverting Eq.~(\ref{eq:Snormal}) one can compute any two-point correlation function of the cavity and the spin. This method is formally equivalent to the quantum regression theorem for Markovian baths: the convenient matrix notation easily extends to the case of several variables.

One specific quantity that can be computed using this method is the number of photons in the cavity $n=\av{a\yd a}$, which is related to the Keldysh Green's function by $2n+1=\int d\omega/(2\pi) G^K(\omega)$. This quantity diverges at the phase transition as\mycite{nagy11,oeztop11}
\be
\av{a\yd a} = \frac{\lambda^2}{2\omega_z\omega_c(1-(\lambda/\lambda_c)^2)}\sim \frac1{\lambda_c-\lambda}
\ee
where here $\lambda_c = (1/2)\sqrt{\omega_z(\omega_c^2+\kappa^2)/\omega_c}$.
{Thus, for the driven-dissipative system, the critical exponent is $\SuscExponent=1$, as in the equilibrium case at finite temperature. This correspondence holds for other properties of the phase transition: for example, although the photon-atom entanglement diverges at the zero temperature transition\mycite{lambert2004entanglement,lambert2005entanglement,wang2014quantum}, this quantity remains finite at the driven-dissipative transition\mycite{wolfe2014certifying}. These observations suggest that the universal properties of driven-dissipative systems are analogous to equilibrium one, at a finite effective temperature. This generic phenomenon will be explained in more detail in Sec.~\ref{sec:LEET}.}

\subsection{Landau theory of a mean-field phase transition}
\label{sec:Landau}
As we have seen, the mean-field critical exponent of the transition at zero temperature differs from the non-equilibrium steady state. This difference can be understood using a simple Landau model of a mean-field Ising transition:
\begin{align} H = \frac{p^2}{2} + \frac12(\lambda_c-\lambda) x^2 + \frac{1}{\jk{4}N} x^4\;.\label{eq:landau}\end{align}
Here $x$ and $p$ are canonical coordinates. This model describes a phase transition at $\lambda_c$: for $\lambda<\lambda_c$ the energy has a single minimum at $x=0$, while for $\lambda>\lambda_c$ two minima are found at \jk{$x_{\rm min}=\pm\sqrt{N(\lambda-\lambda_c)}$}. The effect of spontaneous symmetry breaking corresponds to the choice of one of the two equivalent minima. The expression for $x_{\rm min}$ defines the critical exponent of the model $\OrderExponent=1/2$.
\jk{As discussed in Sec.~\ref{sec:MF}, this matches the equilibrium result for both the zero and finite temperature Dicke model.  The exponent $\OrderExponent$ for the out of equilibrium Dicke model is less straightforward, as it cannot be found from a quadratic theory.  However it is derived in a number of works including\mycite{dimer07} and indeed found to be $\OrderExponent=1/2$.
Thus the Landau theory recovers
 } the correct expression for the Dicke model both at equilibrium and out of equilibrium (see Table \ref{table:critical}).

%The thermodynamic limit of this transition is obtained at $N\to\infty$ and occurs at the critical coupling $\lambda=\lambda_c$. For $\lambda<\lambda_c$, the $N\to\infty$ limit simply corresponds to a regular harmonic oscillator. In contrast, for $\lambda>\lambda_c$, the limit $N\to\infty$ would correspond to an inverted harmonic oscillator. This model is energetically unstable and does not have a well defined equilibrium state, because the partition function $Z=e^{-H/T}$ is not normalizable. To solve this problem, it is necessary to {\it first} expand the Hamiltonian around the global minima, and then consider the thermodynamic limit. The minimal energy is found at $x=\pm\sqrt{N/(\lambda-\lambda_c)}$. This expression determines the order parameter of the phase transition and exemplifies the effect of symmetry breaking: the original model is symmetric under the $Z_2$ transformation $x\to-x$, while the system does not. This derivation is valid for both systems at equilibrium and out-of-equilibrium and defines the critical exponent of the model $\OrderExponent=1/2$.

%A second critical exponent can be defined by considering the divergence of the fluctuations close to the phase transition:
%\be \av{x^2}-(\av{x})^2 = \left\lbrace \begin{array}{c c}A_-(\lambda_c-\lambda)^{\SuscExponent_-} & \lambda<\lambda_c\\A_+(\lambda-\lambda_c)^{\SuscExponent_+} & \lambda>\lambda_c\\\end{array}\right.\ee
%\be \av{x^2}-(\av{x})^2 = A|\lambda_c-\lambda|^{\SuscExponent}
%\ee

The critical exponent $\SuscExponent$ depends on the specific context of the transition. To understand this difference it is sufficient to consider three specific examples of the harmonic oscillator (for simplicity we focus on the normal phase at $\lambda<\lambda_c$):
%
%For simplicity on the case of $\lambda<\lambda_c$ (where $\av{x}=0$): the extension to the case $\lambda>\lambda_c$ is straightforward and delivers the same critical exponent.
%
%

{\it 1. Quantum phase transition (QPT) --} If the system is at zero temperature, $\av{x^2}$ is given by the zero-point motion of the harmonic oscillator, Eq.~(\ref{eq:landau}) with $N\to\infty$, %$\av{x^2}=1/(2\omega_c)$. In the present case $\omega_c^2 =\lambda_c-\lambda$ and
\begin{align}
\av{x^2}_{QPT}=\frac{1}{2(\lambda_c-\lambda)^{1/2}}\;,
\end{align}
leading to the critical exponent $\SuscExponent_{QPT}=1/2$.

{\it 2. Classical phase transition (CPT) --} If the system is at finite temperature, one can apply the equipartition theorem to establish that in the classical limit when $k_B T \gg \ \sqrt{\lambda_c-\lambda}$, then $\av{(\lambda_c-\lambda)x^2}=k_B T$. Thus,
\begin{align}
\av{x^2}_{CPT}=\frac{k_B T}{\lambda_c-\lambda}\label{eq:CPT}
\end{align}
or equivalently $\SuscExponent=1$.  As the mode frequency goes to zero at the transition, the transition point is always in the classical limit,
$k_BT \gg \sqrt{\lambda_c - \lambda}$.

This result also holds for an open system coupled to an equilibrium bath at temperature $T$. In this case the dynamics are described by the Langevin equation
\begin{align}
\ddot x - \eta \dot x + (\lambda-\lambda_c) x^2 = f(t).
\label{eq:langevin}
\end{align}
Here correlations of the Langevin noise $f(t)$ are determined by the fluctuation-dissipation theorem (FDT), $\av{f(t)f(t')}=4\eta k_B T \delta(t-t')$. By inverting Eq.~(\ref{eq:langevin}) one retrieves Eq.~(\ref{eq:CPT})\mycite{van1992stochastic}. As expected, for a classical system the insertion of an equilibrium bath does not modify the (equal-time) correlation functions of the system, and the critical exponent $\SuscExponent$ is left unchanged.

 {\it 3. Non-equilibrium steady state (NESS) --} In the presence of an external drive, the equilibrium FDT is violated, and the random noise source of Eq.~(\ref{eq:langevin}) will be determined by a generic function $\av{f(t)f(t')}=F(t-t')$ and
\begin{align}
\av{x^2}_{\rm NESS} = \int \frac{d\omega}{2\pi}\frac{F(\omega)}{(\omega^2+\lambda-\lambda_c)^2 + \omega^2\eta^2},
\end{align}
where $F(\omega)$ is the Fourier transform of $F(t-t')$. To extract the critical exponent of the transition, it is then sufficient to assume that $F$ is analytic around $\omega=0$, such that for small $\omega$, $F(\omega)\approx F_0$. Under these conditions, for $\lambda\lesssim\lambda_c$,
\begin{align}\av{x^2} \approx \frac{F_0}{2\eta(\lambda-\lambda_c)}\end{align}
and $\SuscExponent_{\rm NESS}=\SuscExponent_{\rm CPT}=1$.

The model (\ref{eq:landau}) allows us to compute a third critical exponent, $\FiniteSizeExponent$. This exponent is defined by the divergence of $\av{x^2}$ at the critical point, $\lambda=\lambda_c$, as a function of $N$. At the transition, the system is governed by $H=p^2/2+(1/N)x^4$. We again need to distinguish the quantum case from the classical one. At zero temperature, the system is found in the ground state of the Hamiltonian, where $\av{p^2}=(1/\jk{4}N)\av{x^4}\sim (1/N)(\av{x^2})^2$. Considering that $\av{x^2 p^2}\sim 1$, one obtains that $\av{x^2}=N^{1/3}$, or $\FiniteSizeExponent=1/3 $. In contrast, at finite temperatures, one can again apply the equipartition theorem to deduce that $\av{x^4/N}=k_B T$, and thus $\av{x^2}\sim N^{1/2}$, or $\FiniteSizeExponent=1/2$.

\begin{table}
\be
\begin{array}{|c |c |c| c| c|}
\hline
{\rm exponent}&{\rm definition}&{\rm QPT}&{\rm CPT}&{\rm NESS}\\
&&~\text{\mycite{vidal2006finite,chen2008numerically,liu09,liu2017universal}}& &~\text{\mycite{oeztop11,nagy11,dalladiehl}}\\
\hline
\OrderExponent & \av{x}\sim \jk{\delta \lambda^{\OrderExponent}} & 1/2 & 1/2 & 1/2\\
\hline
\SuscExponent & \av{x^2-\av{x}^2} \sim |\delta\lambda|^{-\SuscExponent}& 1/2 & 1 & 1\\
\hline
\FiniteSizeExponent&\av{x^2}_{\lambda=\lambda_c} \sim N^{\FiniteSizeExponent} & 1/3 & 1/2 & 1/2\\
\hline
\end{array}\nonumber
\ee
\caption{Critical exponents of mean-field phase transitions (such as the Dicke model). The transition point is set at $\delta\lambda\equiv\lambda-\lambda_c=0$. The critical theory of the non-equilibrium steady state (NESS) are the same as the classical phase transition (CPT) exponents.
}\label{table:critical}
\end{table}

\subsection{Effective low-frequency temperature}
\label{sec:LEET}
Given the equivalence seen above between the thermal and non-equilibrium critical behavior, it is useful to push this connection further and try to identify an effective temperature for the non-equilibrium case. In quantum optics, this is usually done by comparing the mode occupation with an equilibrium ensemble. In the case of the Dicke model, this approach would lead to an effective temperature that diverges at the transition. To describe the critical properties of the transition it is therefore more convenient to focus on the universal low energy behavior, leading to the definition of a low-energy effective temperature (LEET)\mycite{dalladiehl}. The concept of LEET can be understood by considering a single oscillator $x$. The commutation and anti-commutation relations of $x$ at different times are respectively described by
\begin{align}
G^R(t-t') &= i\left[x(t),x(t')\right],&
G^K(t-t') &= i\left\{x(t),x(t')\right\}
\end{align}
The universal properties of the phase transition are determined by the low-frequency expansions of $G^R$ and $G^K$
\begin{align}
{\rm Im}G^R(\omega) &= B \omega + O(\omega^3) &{\rm and}&& G^K(\omega) &= A + O(\omega^2).
\label{eq:Taylor}
\end{align}
Here, we have assumed that both functions are analytic around $\omega=0$ and noted that by definition, they are respectively antisymmetric and symmetric with respect to $\omega\to-\omega$.

The LEET is defined by inspection of the fluctuation-response ratio:
\be
\chi(\omega) \equiv \frac{G^K(\omega)}{{\rm Im}[ G^R(\omega)]}
\ee
At thermal equilibrium $\chi(\omega)={\rm coth}(\omega/2T)$ and in particular at small $\omega$, $\chi(\omega) \approx 2T/\omega$, i.e. a Rayleigh-Jeans distribution. For systems out of thermal equilibrium, $\chi(\omega)$ is a generically unknown function. However, by using Eq.~(\ref{eq:Taylor}), we find that in general
\be
\chi(\omega) \approx \frac{A}{B \omega}\label{eq:Fomega2}
\ee
This expression allows us to define an effective low-frequency temperature as $T^*=A/2B$. Note that in this derivation our only assumption was that $G^R(\omega)$ and $G^K(\omega)$ are analytic around $\omega=0$. For generic non-equilibrium systems, this assumption seems to be valid: the only known exception are quantum systems at zero temperature, where $\chi(\omega)={\rm sign}(\omega)$.

The emergence of a low-frequency effective temperature is a generic feature of non-integrable non-equilibrium systems, and as such has a long history, see e.g.\ Refs.~\cite{hohenberg89,cross93}.  In the context of many-body quantum systems it is predicted to occur in systems as different as voltage-biased two-dimensional gases\mycite{MitraMillis,MitraMillis2,YBKim},  noise-driven resistively shunted Josephson junctions\mycite{us-nature,us-prb}, and BECs of exciton polaritons\mycite{sieberer2013dynamical,sieberer2014nonequilibrium}. This effect has a close analogy to the eigenstate thermalization hypothesis (ETH)\mycite{deutsch1991quantum,srednicki1994chaos,rigol2008thermalization}. This principle states that closed systems generically tend to thermalize at long times. Here, the long time delay after the quench is substituted by low-frequencies, i.e. long time differences between two times in a steady state.

\section{Beyond-mean-field methods}
\label{sec:beyond}
The above-mentioned mean-field analysis has two main limitations: (i) it is valid only in the limit of $N\to\infty$ and (ii) it assumes that all the atoms are coupled homogeneously to the cavity. To overcome these two limitations, different methods have been developed.

\subsection{Bosonic diagrammatic expansion}
As we discussed in Sec.~\ref{sec:HP}, the superradiant transition can be described in terms of Holstein--Primakoff (HP) bosons. Keldysh diagrams offer a natural platform to study $1/N$ corrections, by considering higher order terms in the HP expansion\mycite{dalladiehl,lang2016critical}. Let us, for example, consider the number of photons at the critical coupling, for the driven dissipative model. As discussed in Sec.~\ref{sec:Landau}, this number grows as $N^{1/2}$. The prefactor was computed in Ref.~\cite{dalladiehl} and found to be in excellent agreement with the numerics for small $N$ -- see Fig.~\ref{fig:dalladiehl}. See also Ref.~\cite{lang2016critical} for a study of the relaxation dynamics close to the superradiant transition.

\begin{figure}
\includegraphics[trim={6.4cm 0.5cm 0 0.5cm},clip,scale=0.8]{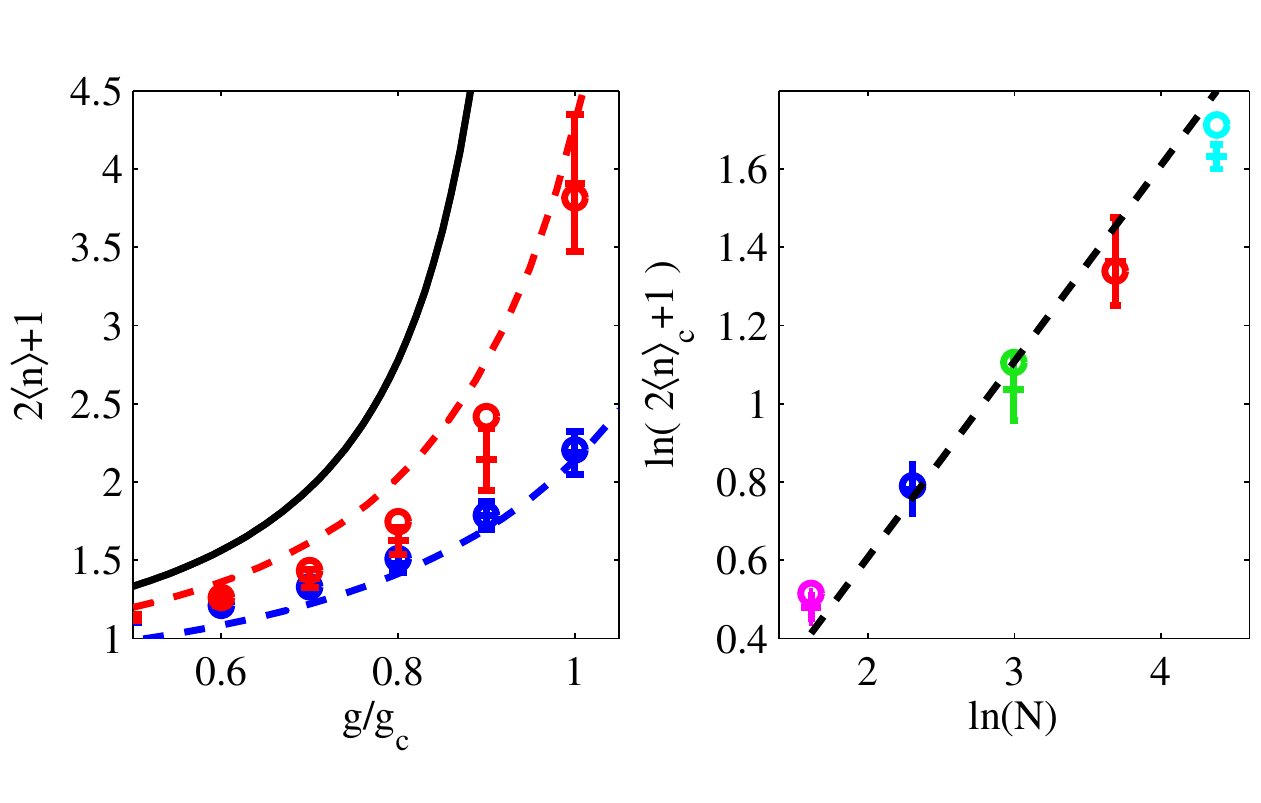}
\caption{Number of photons at the critical coupling $\lambda=\lambda_c$, as a function of $N$, for a Dicke model with $\omega_z=2$, $\omega_c=\kappa=1$, and $\gamma=0$. Diagrammatic expansion (o), Monte-Carlo-wave-function method\mycite{molmer93,vukics2012adequacy} (+), and $\FiniteSizeExponent=1/2$ critical scaling (dashed lines).   Reproduced from Ref.~\cite{dalladiehl}.}
\label{fig:dalladiehl}
\end{figure}

\subsection{Fermionic diagrammatic expansion}

An alternative method to obtain a controlled perturbative expansion in $1/N$ is given by the fermionic path integral approach\mycite{dalla2016dicke}. The key idea is to describe each atomic degree of freedom using the Majorana fermion representation of spin-1/2\mycite{tsvelik2007quantum,shnirman2003spin,schad2015majorana}. In this language the spin is replaced by a complex fermion $f$ and a Majorana fermion $\eta$. The former keeps track of the polarization of the spins $f^\dagger f = 1/2-\sigma_z$, while the latter ensures the correct commutation relations are respected. This formalism allowed the authors of Ref.~\cite{dalla2016dicke} to develop a controlled $1/N$ expansion of the Dicke model. The key result was that to leading order in $1/N$, only one-loop diagrams (and their products) survive. These diagrams can be exactly resummed using the common Dyson resummation, i.e.\ by adding a self-energy contribution to the free Green's function of the cavity:
$[G_a^{R}]^{-1} \to [G_a^{R}]^{-1} +\Sigma_a^R $. Here $[G^R_a]^{-1}$ is the $2\times 2$ upper-left block of Eq.~(\ref{eq:GR_HP}) and $\Sigma_a^R$ is a loop integral. Importantly, this expression simply corresponds to the spin-spin correlation function and can be written as
\begin{align}
\Sigma_a^R(\omega)=&-\frac{8\lambda^2}N \sum_{j=1}^N\int_0^\infty dt~{\rm{Im}}\left[\av{\sigma^x_j(t)
\sigma^x_j(0)}\right] e^{i\omega t}\;.
 \label{eq:sigmaR}
\end{align}
This result has a simple physical meaning: The coupling between the atoms and the cavity is proportional to $1/\sqrt{N}$. Thus, in the limit $N\to\infty$ the {\it feedback} of the cavity onto the atoms is negligible {below threshold}. As a consequence, the cavity feels the {\it free} evolution of the spins, {and the superradiance transition is determined by a sum over $N$ independent terms. This result is analogous to the Lamb theory of lasing\mycite{lamb1964theory,agarwal1990steady,scully97,gartner2011two}, where the feedback of the cavity on the atoms is neglected (see Sec.~\ref{sec:lasing} for a discussion on the similarities and differences between superradiance and lasing)}.

{The superradiant transition occurs when the {\it dressed} Green's function has a pole at zero frequency, or}
\begin{align}
{\rm det}\left[[G_a^R]^{-1}(0)+\Sigma_a^R(0)\right] = 0\end{align}
Substituting Eq.~(\ref{eq:sigmaR}) in the expression for $[G_a^{R}]^{-1}$, we obtain the condition for the superradiant transition  \begin{align}
 &{\rm det}\left[\left(\ba{c c}i\kappa-\omega_c+\Sigma_a^R(0) & \Sigma_a^R(0) \\ \Sigma_a^R(0) & -i\kappa-\omega_c+\Sigma_a^R(0)\ea\right)\right] = 0\nonumber\;,
 \end{align}
where we used the fact that $\Sigma^R_a(0)$ is real by definition. A direct evaluation leads to
\begin{align} \omega^2_c+\kappa^2+2\omega_c\Sigma^R_a(0) = 0\label{eq:gc_sigma}\;,
\end{align}
%
%\texttt{This looks like it assumes $\sigma^R$ is real; should state this.}
%Eq.~(\ref{eq:gc_sigma}) can be combined with Eq.~(\ref{eq:sigmaR}) to compute the critical coupling in the presence of inhomogeneities and losses.

This approach has two limiting cases that coincide with earlier results: (i) For a system at thermal equilibrium $\av{\sigma^z}=(1/2)\tanh(\omega_z/2T)$ and $\gamma=0$. In this case,
$\Sigma_a^R(0)=4\lambda^2\av{\sigma^z}/\omega_z$, and we recover the equilibrium result, Eq.~(\ref{eq:gc_MF}). (ii) In the presence of single-atom decay and dephasing
\be \av{\sigma^x_j(t)\sigma^x_j(0)} = e^{-\gamma_T t}\left[ \cos(\omega_z t) + i\av{\sigma^z} \sin(\omega_z t)\right]\label{eq:corr}\;.\ee
where $\gamma_T=\gamma_\phi+\gamma_\downarrow$. In this case,
$\Sigma_a^R(0) = 4\lambda^2\av{\sigma^z_j}\omega_z/(\omega_z^2+\gamma^2)$, and Eq.~(\ref{eq:gc_sigma}) becomes equivalent to Eq.~(\ref{eq:gc_phi}).

In addition, the present diagrammatic approach allows us to consider inhomogeneous systems: Eq.~(\ref{eq:gc_sigma}) shows that the transition is governed by the disorder-averaged value of $\bar{\lambda}^2 = (1/N)\sum_j \lambda_j^2$. One particular application is the case of inhomogeneous broadening when coupling to Raman transitions between hyperfine states, discussed by Ref.~\cite{zhiqiang2018dicke}.  Furthermore, if the energy splitting of the two-level atoms is disordered, one sees this approach gives the $(\omega_c^2 + \kappa^2)/\omega_c = 4 \langle {\lambda_i^2}/\omega_{z,i} \rangle$. An application of this occurs when considering transitions between motional states of a thermal gas\mycite{piazza2013bose}, for which the two-level system energy, $\omega_{zi} = \epsilon_{\vec{k}_i +\vec{Q}_\text{recoil}} - \epsilon_{\vec{k}_i}$ with $\epsilon_{\vec{k}} = \hbar^2 k^2/2m$, depends on the Boltzmann distributed initial momentum of the atoms.

\subsection{Cumulant expansion}

\label{sec:cumulants}
A further way to consider systems with finite $N$ is to derive a hierarchy of coupled equations for all moments of the photon and spin operators. In the thermodynamic limit, $N\to\infty$, only the mean-field parts of these equations survive while at large but finite $N$ the second order correlation functions can give an accurate picture of the behavior.

When analyzing the dynamics using simply mean-field theory it is necessary to introduce symmetry breaking terms by hand. This is because the normal state is always a solution to the mean-field equations. By considering the second moments of the distribution one may look for discontinuities in quantities such as the photon number which respect the $\mathbb{Z}_2$ symmetry of the model.  This allows us to only consider a reduced set of equations for the second moments which respect these symmetries. These techniques are closely related to those used in laser theory to describe the emergence of spontaneous coherence there\mycite{Haken1970,Haken1975}.

For the Dicke model there are three distinct classes of these equations. The first are those that describe correlations of the photon mode
\begin{align}
  \partial_t\average{\adag a} &= -2\kappa\average{\adag a} - \lambda N\Im [\ax]  \\
  \partial_t\average{aa} &= -2(i\omega_c+\kappa)  \average{aa} - i\lambda N\ax
\end{align}
where we have denoted $\ax = \exax$. The second type of equations are those which involve correlations between the photon and spin degrees of freedom:
\begin{gather}
  \begin{split}
  \partial_t\ax =
  -\left(i\omega_c+\kappa+\gamma_T\right)\ax - \omega_z \ay
  \\ -i\lambda\left[\left(N-1\right)\xx +\frac12 \right],
  \end{split}
  \\
  \begin{split}
  \partial_t\ay =
  -\left(i\omega_c+\kappa+\gamma_T\right)\ay  - \lambda\sz\left(\aa+\ada\right)
  \\
    + \omega_z\ax - i\lambda\left[(N-1)\xy-i\frac12\sz\right].
  \end{split}
\end{gather}
In these equations $\abx$ means $\langle \sigma^\alpha_i \sigma^\beta_{j \neq i}\rangle$ the correlation between $\sigma^\alpha$ at one site and
$\sigma^\beta$ at another.  All such correlations are equivalent since each atom is identical.  These cross correlations
obey:
\begin{align}
  \partial_t\xx &= -2\omega_z\xy - 2\gamma_T\xx, \\
  \partial_t\yy &= 2\omega_z\xy -2\gamma_T\yy - 4\lambda\sz\Re[\ay],
  \\
  \partial_t\zz &= 4\lambda\sz\Re[\ay] -4\gdn\left(\zz +\frac12\sz\right),
  \\
  \partial_t\xy &= \omega_z(\xx-\yy)-2\gamma_T\xy -
  2\lambda\sz\Re[\ax].
\end{align}
In writing these expression we have broken third order moments into products of first and second moments by assuming that the third order cumulants vanish. These equations do not put any restrictions on the types of decay processes which can be present and those written above include both collective decay channels such as photon loss and individual atomic loss and dephasing.

{In most cases, the decay channels only shift the position of the transition. One important exception was found by Ref.~\cite{kirton2017suppressing}, who showed that the presence of dephasing ($\gamma_\phi$) without losses ($\gdn=0$) completely suppresses the transition: This effect is demonstrated in Fig.~\ref{fig:kirton}, which {shows the behavior at} a value of the coupling far above the mean-field prediction for the location of the transition. This figure shows the reduced photon number ($\av{a^\dagger a}/N$), as a function of $N$, for various combinations of loss processes.  In the case of $\gdn=0$, the dynamics always reaches a normal state with an average photon number that scales only as $\sqrt{N}$. This effect is due to the depolarization of the atoms due to sub-leading terms in the $1/N$ expansion, which can be compensated by decay processes ($\gdn\neq0$) that polarize the atoms. As we will see below, this prediction is in good agreement with the numerical results obtained for finite $N$.}

\begin{figure}
\includegraphics[clip,scale=0.8,scale=0.4]{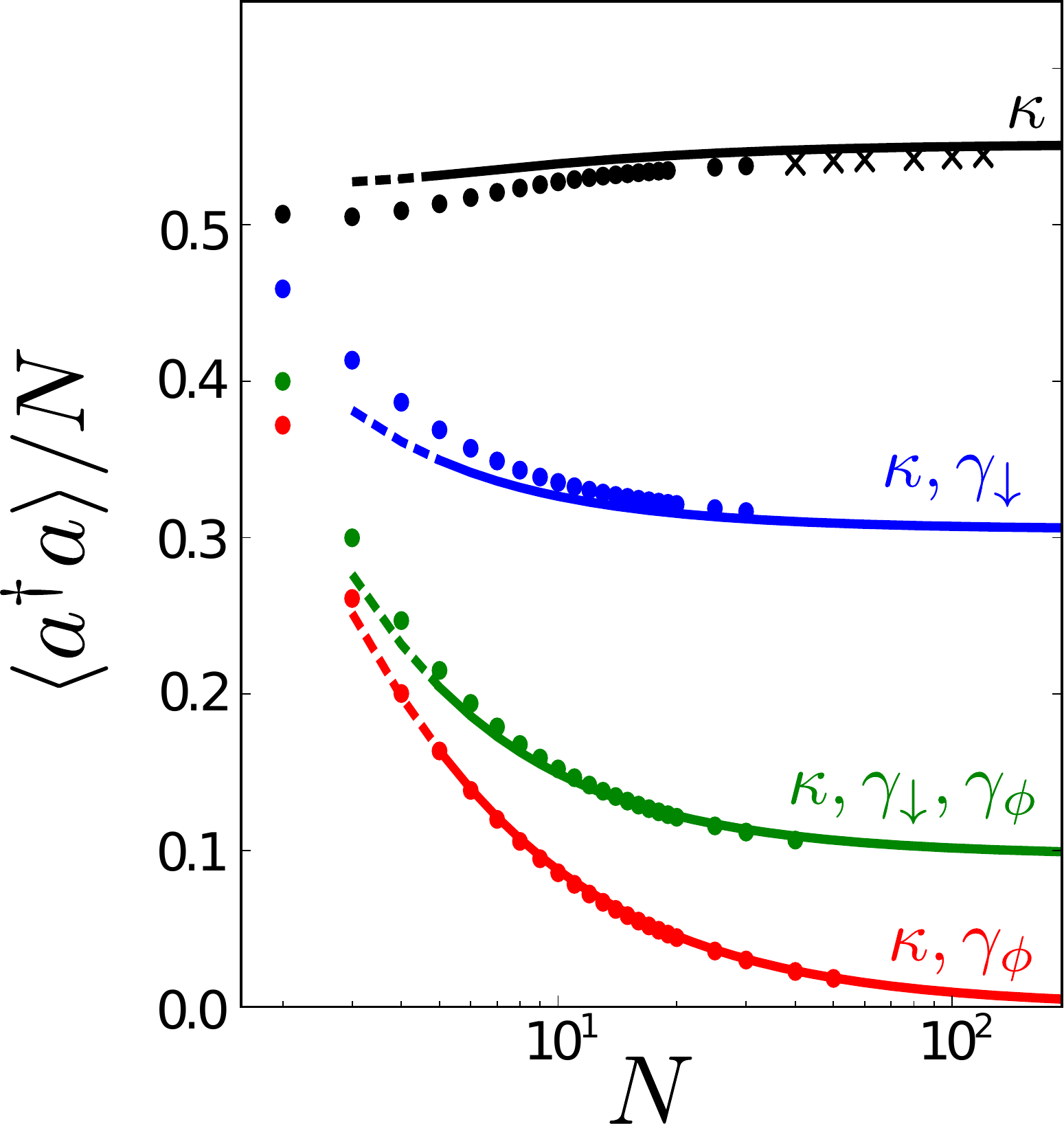}
\caption{Number of photons for $\lambda>\lambda_c$, obtained from the cumulant expansion method (lines) and from the numerics (dots). The lines correspond, from top to bottom, to
   $\gdn=\gphi=0$ (black), $\gdn=0.1$, $\gphi=0$ (blue), $\gdn=0.1$,
   $\gphi=0.2$ (green) and $\gdn=0$, $\gphi=0.02$ (red). Other parameters are: $\lambda\sqrt{N}=0.9$, $\omega_c=1$,
   $\omega_z=1$, $\kappa=1/2$. Reproduced from Ref.~\cite{kirton2017suppressing}.}
   \label{fig:kirton}
\end{figure}

\subsection{Numerical approaches}

For small numbers of atoms it is straightforward to find the exact Hamiltonian or Liouvillian of the appropriate model,  determine the density operators in a thermal or steady-state ensemble, and calculate all possible observables.
To reach larger system sizes it is possible to use the collective spin representation of the Dicke model as in Eq.~\eqref{eq:H_totalS}. The Hilbert space dimension then scales linearly with the number of atoms and so the
problem can again be straightforwardly diagonalized. This approach is, however, limited to only studying collective decay processes. More sophisticated methods are required to study the problem efficiently when individual loss processes are present.

In this more general case, a subtle symmetry can be exploited to efficiently calculate the behavior of the system. This remaining symmetry is a \textit{permutation symmetry} at the level of the density matrix rather than in the Hilbert space: If the master equation can be written as a sum of
processes where each term only affects a single site $i$, then swapping any pair of sites leaves the state unchanged. In this case, each element of the density matrix (ignoring the photon) must obey:
\begin{align}
&\bra{s^L_1 \ldots s^L_i \ldots s_{j}^L \ldots s_N^L}\rho\ket{s^R_1 \ldots s^R_i \ldots s_j^R \ldots s_N^R}\nonumber \\
&\equiv \bra{s^L_1 \ldots s^L_{j} \ldots s_{i}^L \ldots s_N^L}\rho\ket{s^R_1 \ldots s^R_{j} \ldots s_{i}^R \ldots s_N^R},\nonumber
\end{align}
where $s^{L(R)}=\pm1/2$. The full density matrix then separates into sets of permutation-symmetric elements.
To find the dynamics of the system it is sufficient to propagate a single representative element from each of these sets, therefore gaining a combinatoric reduction to the size of the Liouvillian.
 The steady state can also be calculated by finding, in this restricted space, the eigenvector of the Liouvillian with eigenvalue 0.

This approach has been applied to a variety of problems which preserve this permutation symmetry. For example, it was used to study spin ensembles\mycite{Chase08}, lasing models\mycite{Xu13}, coherent surface plasmons\mycite{Richter2015}, the competition between collective and individual decay channels\mycite{Damanet2016}, the behavior of an ensemble of Rydberg polaritons\mycite{Gong2016}, equilibrium properties of a model with a larger local Hilbert space\mycite{Zeb2018},  subradiant states in the Dicke model\mycite{Gegg2017a}, the effect of individual losses on transient superradiant emission\mycite{Shammah2017} and the crossover between superradiance and lasing\mycite{Kirton2018} (see Sec.~\ref{sec:generalized}). {These results are reviewed in Ref.~\cite{shammah2018}}, while libraries which implement this method can be found at Refs.~\cite{Kirton2017a, Gegg2017, shammah2018piqs}.

This method was also applied to the Dicke model, to study the effect of individual loss processes on the superradiant transition. As shown in Fig.~\ref{fig:kirton}, the numerical results are in {\it quantitative} agreement with the above-mentioned cumulant expansion\mycite{kirton2017suppressing}, valid for large $N$.
%We see that at the largest values of $N$ reachable by the numerics there is good agreement with the cumulant expansion.
Thus, a combination of these two methods is able to cover the entire range of number of atoms; from $N=1$ to $\infty$.}

\section{Superradiance and lasing}
\label{sec:lasing}

 \jk{In this section we discuss the relation between the superradiance transition and lasing.  To make this connection clear\edt{, in Sec.~\ref{sec:TC} we first discuss a canonical model of lasing, namely the Tavis--Cummings model. Next, in Sec.~\ref{sec:generalized} we introduce a} generalized Dicke model that interpolates between the Dicke and the Tavis--Cummings model. This family of models provides a link between the superradiant transition and the closely related phenomenon of lasing.}
 In Secs.~\ref{sec:counter}-\ref{sec:superlaser}, we describe different types of lasing transitions (regular lasing, counter lasing, and superradiant lasing) and explain their similarities and differences with the superradiant transition.

\subsection{The Tavis--Cummings model}
\label{sec:TC}
The Tavis--Cummings model is given by a Dicke model without counter-rotating terms:
\begin{align}
H = \omega_c a^\dagger a + \omega_z\sum_{j=1}^N \sigma^z_j + \frac{\lambda}{\sqrt{N}}\sum_{j=1}^N (a \sigma^+_j + a \sigma^-_j) \label{eq:TC}
\end{align}
This model conserves the total number of excitations $N_{\text{ex}}=a^\dagger a + \sum_j \sigma^z_j$. This symmetry is associated with a $U(1)$ gauge symmetry $a\to e^{i\phi}a$ and $\sigma^-\to e^{i\phi}\sigma^-$. The equilibrium Tavis--Cummings model has a phase transition at $\lambda=\sqrt{\omega_c\omega_z}$, where the symmetry is spontaneously broken. This critical coupling differs by a factor of two from the Dicke result, as only half the matter-light coupling terms are present.

In the presence of decay, the Tavis--Cummings model does not show a superradiant transition\mycite{Keeling2010,larson2017some,soriente2018dissipation}. This result has a simple physical meaning: because the model does not have counter-rotating terms, it will always flow to a trivial steady state, where the cavity is empty and the spins are polarized in the $\sigma^z=-1/2$ direction. The superradiant transition occurs only if the total number of excitations is kept constant ({when no loss processes are present}). The Tavis--Cummings model can nevertheless show a {\it lasing} transition if the atoms are pumped. In what follows, we explain the difference between the superradiant transition and the lasing transition, by considering a simple model in which both transitions occur.

\subsection{Generalized Dicke model}
\label{sec:generalized}
The generalized Dicke model is a simple interpolation between the Dicke model (\ref{eq:Dicke}) and the Tavis--Cummings model (\ref{eq:TC}),
\begin{align} H= \omega_c a\yd a+\omega_z\sum_{j=1}^N\sigma_j^z&+\frac{\lambda}{\sqrt{N}}\sum_{j=1}^N(a \sigma^+_j +a\yd \sigma^-_j)
\nonumber\\& +  \frac{\lambda'}{\sqrt{N}}\sum_{j=1}^N (a \sigma^-_j +a\yd \sigma^+_j)\;.\label{eq:genDicke}\end{align}
This model includes the Dicke model ($\lambda=\lambda'$) and the Tavis--Cummings model ($\lambda'=0$) as special cases. It can be  realized using the 4-level scheme described in Sec.~\ref{sec:models}, where rotating and counter-rotating terms are induced by two separate pumping fields.

Using the Holstein--Primakoff approximation\mycite{holstein40}, one can map this model to two coupled harmonic oscillators:
\be H=\omega_c a\yd a+\omega_z b\yd b + \lambda (a b\yd + a\yd b) +  \lambda'(a b + a\yd b\yd)\;. \label{eq:Dicke_HP}\ee
This Hamiltonian can be represented as a $4\times4$ matrix
\be H = \frac12(a~ a\yd~ b~ b\yd)\left(\ba{c c c c}\omega_c & 0 & \lambda & \lambda' \\0 & \omega_c & \lambda' & \lambda\\ \lambda & \lambda' & \omega_z & 0 \\ \lambda' & \lambda & 0 & \omega_z \ea\right)\left(\ba{c}a\yd \\ a\\b\yd\\ b\ea\right)
\ee
where $\lambda_\pm=\lambda\pm \lambda'$. Following the same analysis as in Sec.~\ref{sec:HP} one obtains
%The corresponding action is simply  obtained by the Legendre transformation $S = \int dt \left[~i~ a\yd\partial_t a - H\right] = \int d\omega~(-\omega) a\yd(\omega)~a(\omega) - H$, leading to the inverse retarded Green's function
\be G_R^{-1} = \left(\ba{c c c c}\omega-\omega_c+i\kappa & 0 & -\lambda & -\lambda' \\0 & -\omega-\omega_c-i\kappa & -\lambda' & -\lambda\\-\lambda & -\lambda' & \omega-\omega_z & 0 \\ -\lambda' & -\lambda & 0 & -\omega-\omega_z \ea\right)
\label{eq:GR_generalized}
\ee
where $\kappa$ is the cavity decay rate. The superradiant transition is signaled by ${\rm det}[G_R^{-1}(\omega=0)]=0$, or
\be
(\lambda^2-\lambda'^2)^2 - 2(\lambda^2+\lambda'^2)\omega_c\omega_z + (\kappa^2+\omega_c^2)\omega_z^2 = 0
\label{eq:gc_generalized}
\ee
Let us now consider the two above-mentioned limiting cases: in the Dicke model ($\lambda=\lambda'$), one recovers Eq.~(\ref{eq:gc_collective}). In contrast, for the Tavis--Cummings model ($\lambda'=0$). the superradiant transition occurs for
\be
(\lambda^2 - \omega_c\omega_z)^2 + \kappa^2\omega_z^2 = 0.
\ee
This condition cannot be satisfied for any $\kappa\neq0$, in agreement with the results of Sec.~\ref{sec:TC}. In general, for any finite $\kappa$, the critical coupling diverges when approaching the TC limit of $\lambda'\to 0$\mycite{Keeling2010}.
\jk{
It is worth also noting that identical behavior occurs if we set $\lambda=0$ and consider the model with only \emph{only} counter-rotating terms.  In fact this limit is also the Tavis-Cummings model after a unitary transform, rotating the spin by $\pi$ about the $x$ axis, thus sending $\sigma^\pm \to \sigma^\mp$ and $\sigma^z \to - \sigma^z$.

 When considering the full phase diagram of dissipative Dicke model with $\lambda \neq \lambda^\prime$, some new features can arise.  In particular
there exists a phase where both the normal state and superradiant state are stable, and a multicritical point where this phase vanishes, as has been reported a number of times\mycite{Keeling2010,soriente2018dissipation,gutierrez2018dissipative}.
}

\subsection{Regular and counter-lasing transitions}
\label{sec:counter}
Although the Tavis--Cummings model cannot undergo a superradiant transition, this model can describe the transition to a lasing state\mycite{scully97}. 
\jk{We first discuss how this distinct form of coherent light arises in this model, before considering how the lasing state and superradiant states can be related and distinguished.}
To obtain lasing, it is sufficient to supplement the TC model, Eq.~(\ref{eq:TC}), by an incoherent driving term that pumps the atoms in the excited state. This effect can be described by adding a Lindblad operator to Eq.~(\ref{eq:Lindblad}) where $L=\sigma^+$ with a rate $\gamma_\uparrow$.
%Such a term, i.e. incoherent pumping to the excited state really corresponds to a process such as incoherent Raman scattering (i.e. involving Raman scattering into free space), giving a rate of ground to excited state transitions.
This process is directly analogous to a three-level model of a laser, where one of the levels is pumped incoherently, leading to population inversion.  The resulting phase transition leads to a lasing state, rather than a superradiant state. 

\begin{figure}
\includegraphics[scale=0.25]{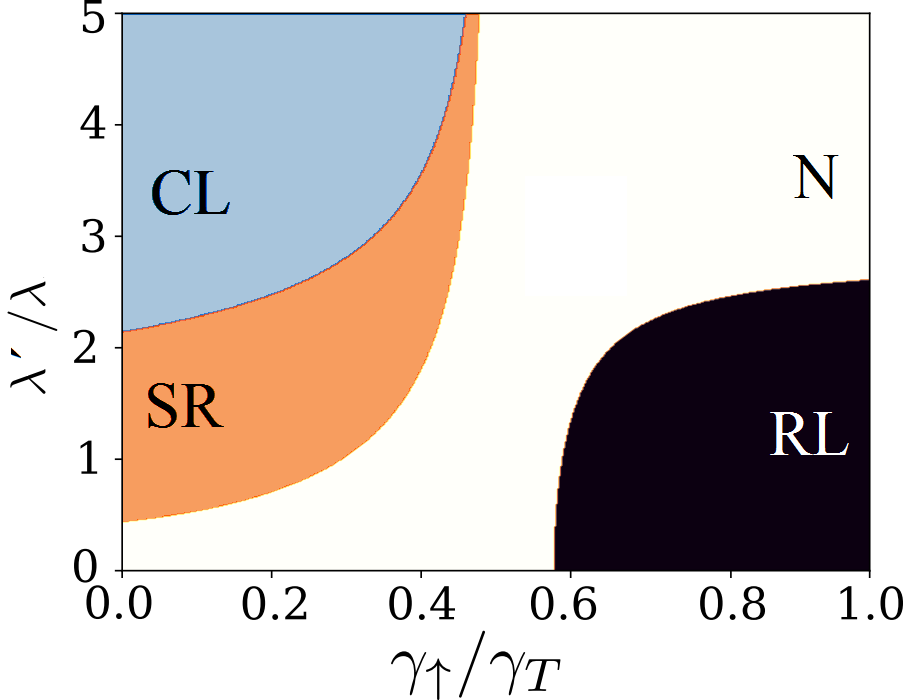}
\caption{Phase diagram of the generalized Dicke model, Eq.~(\ref{eq:genDicke}), with repumping $\gamma_\uparrow$. This model shows regions of superradiance (SR), counter-lasing (CL), and regular lasing (RL). A normal (N) region separates the regions without population inversion (SR and CL) from the regular lasing region. Numerical parameters:  $\omega_z=1,~\omega_0 = 1,~ \lambda=0.9, \kappa=0.5,~\gamma_T = 0.5$. Adapted from Ref.~\cite{Kirton2018}.}
\label{fig:SRlaser}
\end{figure}

\jk{The relation of lasing and superradiance is made clear if one considers the generalized Dicke model (with $\lambda^\prime \neq \lambda$) combined with the incoherent pumping discussed above\mycite{Kirton2018}.  In this case, one sees two distinct ordered states:  \edt{a} lasing state \edt{that} continuously connects to the state with $\lambda^\prime =0$, and a superradiant state that connects to the Dicke model with $\lambda^\prime=\lambda, \gamma_\uparrow=0$.  These two states occupy disconnected regions on the $\lambda^\prime, \gamma_\uparrow$ phase diagram  -- see Fig.~\ref{fig:SRlaser}.}
From a physical perspective, the lasing and superradiant transitions can be clearly distinguished as lasing only occurs when $\av{\sigma^z}>0$, while the superradiant state occurs only for $\av{\sigma^z} < 0$\mycite{Kirton2018}.

{In addition to the presence or absence of inversion, the lasing and superradiant phases have}
%The counter-lasing instability is generically found in the vicinity of the superradiant phase, but has
 a different nature: In the superradiant phase the field is locked to the rotating frame of the pump. In contrast, in a lasing phase, the coherent emission is not locked to the pump frequency and is time dependent in the frame of the pump. From a mathematical perspective the Dicke transition corresponds to a subcritical pitchfork instability, where a single eigenvalue vanishes\mycite{strogatz2018nonlinear}. In contrast, the lasing transition corresponds to a critical Hopf bifurcation, i.e. to a point where two eigenvalues become unstable simultaneously, by crossing the real axis without passing through the origin. Because the unstable modes have a finite real part, this transition generically leads to oscillations. Other examples of Hopf bifurcations in generalized Dicke models were predicted by Ref.~\cite{bhaseen12} and Ref.~\cite{genway2014generalized}, who considered the effects of additional terms, such as $U S_z^2$ and $\Omega S_x$. When the instability is crossed, the system generically gives rise to oscillating superradiant phases, described by limit cycles {\mycite{bhaseen12,piazza2015self}}.

\begin{figure}
\vspace{0.2cm}
\includegraphics[trim={0cm 0cm 0cm  0},clip,scale=0.8,scale=0.48]{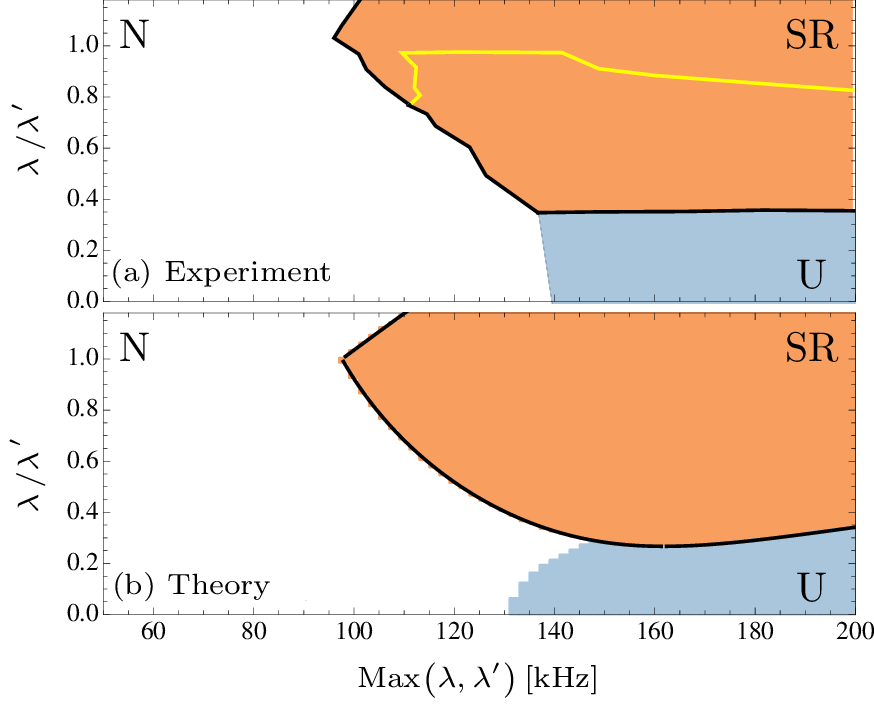}
\caption{Comparison between (a) experimental and (b) theoretical phase diagrams for the generalized Dicke
system, Eq.~(\ref{eq:genDicke}), in the absence of repumping. The system can be either normal (N), superradiant (SR), or unstable/counter-lasing (U). The SR regime below the yellow line shows transient oscillations in time and is possibly related to the oscillating superradiance of Ref.~\cite{bhaseen12}. The parameters used for theoretical calculation correspond to the experimental values: cavity mode frequency $\omega_c = 100$ kHz, dissipation $\kappa = 107$ kHz, and
energy splitting $\omega_z = 77$kHz. The atomic polarization is assumed to  be $\av{\sigma^z}=-0.25$ and the dissipation $\gamma_T= 30 $kHz. Reproduced from Ref.~\cite{shchadilova2018fermionic}.\vspace{-0.5cm}}
\label{fig:schadilova}
\end{figure}

{In addition to standard lasing for the inverted state, a lasing instability  can alternatively be obtained for the Dicke model with negative {detuning of the cavity} ($\omega_c<0$), where the superradiant transition does not occur\mycite{klinder2015dynamical,zhiqiang2018dicke}.}  {Moreover,}
{even in the absence of incoherent pumping ($\av{\sigma^z}<0$) and for positive cavity detunings ($\omega_c>0$)}, a lasing transition can be obtained in the generalized Dicke model of Sec.~\ref{sec:generalized}. This transition occurs when the counter-rotating terms lead to a coherent emission of photons from the cavity. It was termed the ``inverted-lasing''\mycite{Kirton2018} or ``counter-lasing''\mycite{shchadilova2018fermionic} transition and had been observed experimentally by Zhiqiang \textit{et al.}\mycite{zhiqiang2017nonequilibrium}, see Fig.~\ref{fig:schadilova}.

% In addition to the superradiant transition, the generalized Dicke model can undergo a lasing transition. These two transitions correspond to two types of instabilities, or bifurcations. In the former transition a single eignefrequency crosses the origin of the complex plane and becomes unstable. In the latter, two instabilities cross simultaneously the real axis and become unstable. Th
%\texttt{Both transitions are continuous, i.e. second order transitions.  A first order transition in dynamical systems language corresponds to a subcritical Hopf bifurcation, but the lasing transition is critical Hopf.}
%These two types of phase transitions are analogous to second order and first order phase transitions, respectively. Indeed, only the superradiant transition, where an eigenvalue vanishes, is characterized by diverging quantities and critical exponents.

\subsection{Superradiant lasers}
\label{sec:superlaser}
As noted above, the Tavis--Cummings model with incoherent pumping can undergo a transition to a coherent state, i.e.\ lasing.  The connection between this transition and the transient superradiance discussed by Dicke has been considered a number of times\mycite{kolobov93,haake93,meiser09,meiser10}. As mentioned in Sec.~\ref{sec:history}, in the absence of a cavity, transient superradiance produces a coherent pulse by effectively synchronizing the emission of all atoms through the collective decay process.  By placing many atoms in a bad cavity, and continuously incoherently repopulating the excited state, one may try to drive a continuous superradiance process, which has been termed a superradiant laser\mycite{haake93}.  Such a device based on atomic transitions can boast a very narrow linewidth, determined by the sharply defined atomic resonance frequency, rather than the cavity. If one uses a suppressed electronic transition for the lasing level, this allows a very small natural linewidth $\gamma$, but yet superradiant lasing can emerge in the collective strong coupling regime, $N \lambda^2 \gg \kappa \gamma$.   Moreover, the linewidth at peak lasing power scales as $N^{-2}$;  this suggests a potential mHz linewidth from $10^6$ atoms, a level that could significantly improve atomic clock accuracies\mycite{meiser09}.

Earlier works\mycite{haake93} were based on a three-level lasing scheme, and did not address how superradiant lasing arises in the presence of individual decay and dephasing of the atoms.  A simpler two-level description was given in Refs.~\cite{meiser09,meiser10}, using the cumulant expansion approach described in Sec.~\ref{sec:cumulants}.  Such a superradiant laser has been realized experimentally, in a scheme where the lasing transition was actually a two photon Raman transition\mycite{bohnet2012steady,Bohnet2012a}, enabling tuning of both the matter light coupling $\lambda$ and the effective natural linewidth $\gamma$ of the transition.
\vspace{0.5cm}

\jk{
\section{Closely related models}
\label{sec:clos-relat-models}

\pk{So far in this review, we have focused} on the Dicke model, as well as the generalized Dicke model in which we allow distinct strengths of the rotating and counter-rotating terms.  There do however exist a number of models that are closely related to the Dicke model, involving coupling between many two-level systems and a common cavity mode, as well as models such as the Rabi model that can be shown to have a close connection to superradiance.  Here we provide a brief summary of these models, and the novel physics they can introduce.

\subsection{Extended Dicke models}
\label{sec:gener-dicke-models}
The Raman driving scheme \edt{used to realize the Dicke model generates additional terms that need to be taken into account. In particular, the difference between} the cavity-photon-induced Stark shifts in the two atomic states \edt{leads to} a term $U a^\dagger a \sigma^z$. This \edt{term} can also be seen as \edt{a modification of} the cavity frequency depending on the atomic state.  For the motional state reali\edt{z}ation, such a term is inevitable (due to the different overlaps between the two momentum states with the cavity optical lattice)\mycite{nagy10,baumann10}.  For the Raman reali\edt{z}ation, the strength of \edt{$U$} can in principle be turned to zero\mycite{dimer07}. Such a term has been studied extensively in Ref.~\cite{bhaseen12}, where it was seen to enable bistability between normal and superradiant states, as well as distinct superradiant states and time-dependent attractors, i.e.\ limit cycles.

Another additional term that can be easily engineered is a drive, $H_F = F(a + a^\dagger)$, which corresponds to a coherent light source coupled directly to the cavity mode. \edt{This term has the same physical effect as} $H_F^\prime = F \sigma_x$, \pk{these} two forms being \pk{related} by a unitary transformation.  This latter \edt{term} arises naturally in many reali\edt{z}ations of the Dicke model using trapped ions\mycite{porras04,wang13,genway2014generalized,pedernales2015quantum,aedo2018analog}.  \edt{These two terms} break the \edt{$\mathbb{Z}_2$} symmetry \edt{of the Dicke model, and \pk{thus} destroy the} phase transition.  However, there can still be optical bistability\mycite{bowden79} between a high field and low field state, i.e. the open-system analog of a first order phase transition.  The behavior of this model at large driving has also been recently discussed in Ref.~\cite{gutierrez2018dissipative}, establishing the connection to breakdown of the photon blockade seen in the single-atom Jaynes-Cummings model\mycite{carmichael15}.

\subsection{Disordered Dicke model}
\label{sec:disord-dicke-model}

The above models involve adding extra terms to the Dicke model; another class of closely related models involves considering the role of disorder. i.e., returning to Eq.~(\ref{eq:Dicke}) in terms of individual two-level systems, and allowing
different energies or coupling strengths for different systems:
\begin{equation}
H = \omega_c a^\dagger a + \sum_{j=1}^N \omega_z^j \sigma^z_j +\frac{2}{\sqrt{N}} (a+a^\dagger) \sum_j \lambda^j \sigma^x_j\;. \label{eq:Dicke-diss}
\end{equation}
Such models have been studied in a wider variety of contexts, including the effects of disorder on dynamical superradiance in a low Q cavity\mycite{temnov05superradiance},
the phase diagram of microcavity polaritons\mycite{marchetti06thermodyn,marchetti07absorb},  solid state quantum memories\mycite{diniz2011strongly}, as well as for cold atom in optical cavities, accounting for the spatial variation of the cavity modes\mycite{zhiqiang2018dicke}. \edt{Several works} in this context ha\edt{ve} investigated the dynamics of an initially prepared state, using either brute force numerics for small systems\mycite{tsyplyatyev09dynamics,krimer2014non,krimer2016sustained}, or matrix product state approaches\mycite{dhar2018variational}.  The existence of such disorder prevents the simplification of replacing individual spins by a collective spin operator, hence the need for efficient numerical methods to explore this enlarged Hilbert space\mycite{dhar2018variational}.  It is however notable that in the case where $\omega^j_z$ is disordered, while $\lambda^j = \lambda$, the model can be shown to be integrable, as a special case of a Richardson-Gaudin model\mycite{dukelsky04exact,kundu04quantum,tschirhart14,yuzbashyan2018}.  In addition to the dynamics, one can also calculate the phase diagram of the disordered Dicke model by mean-field approaches\mycite{marchetti06thermodyn,marchetti07absorb}, showing \edt{that the} disorder does not destroy the superradiant phase, but modifies the phase boundary.

\subsection{Floquet Dicke models}
\label{sec:floquet-dicke-models}

Another class of driven-dissipative generalized Dicke models involve \pk{time dependent couplings}.  In particular, Floquet-Dicke models where $\lambda(t) = \lambda_0 + \Delta \lambda \cos(\Omega t)$ have been considered\mycite{bastidas12}. \edt{These models show} a complex phase diagram, depending on the ratio of the drive frequency to other energy scales in the model.  Recent work on the same model has studied how time dependent driving can suppress the formation of the superradiant state\mycite{cosme18}.

\subsection{Scaling limit of the Rabi model}
\label{sec:Rabi}

We finally consider a model that has a quite different structure to the Dicke model, but nonetheless can show a similar superradiance transition.  This is the  Rabi model, describing} the coupling between a {quantized {harmonic oscillator}} and a {\it single} spin:
\be
H = \omega_c a^\dagger a + \omega_z \sigma^z + 2\lambda (a+a^\dagger)\sigma^x\;.
\ee
To observe the superradiant transition in this model, Hwang \textit{et al.}\mycite{hwang2015quantum,hwang2018dissipative} proposed considering the limit in which the atomic splitting $\omega_z$ tends to infinity. This limit can be formally studied by defining $\omega_z=\eta \tilde{\omega}_z, \lambda=\tilde{\lambda}\sqrt{\eta}$ and considering the limit of $\eta\to\infty$ such that $\lambda^2/(\omega_c \omega_z)$ remains finite.  If one considers the mean field ansatz of Sec.~\ref{sec:MF}, one finds the ground state free energy
\begin{math}
  F(\alpha)=\omega_c\alpha^2 - \jk{\frac{1}{2}}\sqrt{ \eta^2 \tilde{\omega}_z^2 + \jk{16} \eta \tilde{\lambda}^2 \alpha^2}.
\end{math}
In order to consider the limit $\eta \to \infty$, it is convenient to consider $\alpha = \sqrt{\eta} x$ which gives:
\begin{align}
  F(x)=\eta\left[
    \omega_c x^2 - \jk{\frac{1}{2}}\sqrt{ \tilde{\omega}_z^2 + 16 \tilde{\lambda}^2 x^2}
  \right].
\end{align}
This expression is equivalent to the $T=0$ form of Eq.~(\ref{eq:mft-f}) with
$\eta$ playing the role of the number of atoms.  In the limit $\eta\to\infty$, there is a sharp phase transition at $\tilde{\lambda}=\sqrt{\tilde{\omega}_z \omega_c}/2$, analogous to the Dicke model.

The phase transition of this model can also be found by adiabatically eliminating the state of the two-level system using a \jk{Schrieffer-Wolff transformation}, leading to an effective photon-only problem
\begin{align}
  H = -\jk{\frac{\omega_z}{2}} + \omega_c a^\dagger a -
  \frac{\lambda^2}{\omega_z} (a+a^\dagger)^2
\end{align}
After a Bogoliubov transformation, this expression gives a photon frequency,
$\sqrt{\omega_c(\omega_c - 4\lambda^2/\omega_z)}$, which vanishes at the transition.

% The Rabi model then becomes:
% \be
% H =  \omega_c a^\dagger a + \eta \omega_c \sigma^z + 2\lambda (a+a^\dagger)\sigma^x
% \ee
% For large $\eta$ the spin is highly polarized and can be approximated using the Holstein-Primakoff method. In this limit one obtains
% \be
% H = \eta \left[ \omega_c \alpha^\dagger \alpha + \omega_c \sigma^z + \frac{2}{\sqrt{\eta}}\lambda (a+a^\dagger)\sigma^x \right]
% \ee
% where we defined $\alpha=a/\sqrt{\eta}$. This Hamiltonian is formally equivalent to Eq.(\ref{eq:Dicke_HP}) and gives rise to a superradiant transition. Note that the bosonic operators $\alpha$ and $\alpha^\dagger$ have vanishing commutation relations. However, because the non-equilibrium superradiant transition is governed by classical fluctuations, the absence of quantum fluctuations is actually negligible.

\section{Conclusion}
The Dicke model is one of the fundamental models of cavity quantum electro-dynamics (cavity-QED), describing the coupling of many atoms to a single cavity mode. The thermodynamic limit of this model is achieved by considering an infinite number of atoms, whose coupling to the cavity tends to zero. This model can undergo a phase transition to a superradiant state at a critical value of the light-matter coupling.  \jk{Various physical reali\edt{z}ations of this model have been considered, which may be thought of as analog quantum simulators of the Dicke model, built from driven atoms in cavities, superconducting qubits, or trapped ions.}
%two distinct transitions: a superradiant transition for positive detunings, and a lasing transition for negative detunings.
In this Progress Report, we introduced the reader to the equilibrium and non-equilibrium behavior of this model and showed how to calculate the critical properties of the superradiant transition. For simplicity, we focused on the simplest realization of the Dicke model where mean-field theory gives a good understanding of the behavior. {Our discussion focused on the theoretical aspects of the transition. Experiments were able to probe a diverging susceptibility at the transition\mycite{landig2015measuring}, but the critical exponents were not found to match the theoretical expectations\mycite{brennecke2013real}. This point certainly deserves further investigation.}

A natural generalization of this model involves two coupled cavity modes, leading to a competition between two superradiant phases. At the interface between these two phases the model shows an enlarged $U(1)$ symmetry\mycite{Fan14,Baksic14}, as realized experimentally recently\mycite{Leonard17a,Leonard17b}.  Such experiments have prompted theoretical discussion of the possibility of a vestigial ordered phase\mycite{Gopalakrishnan17}, where the two cavities become phase locked but without superradiance, as well as the nature of the excitations close to the $U(1)$ symmetric point\mycite{lang2017collective}. A further extension in this direction leads to multi-mode cavities, which give rise to spatially varying, cavity-mediated interactions among the atoms
\mycite{kollar15,Kollar2016,vaidya18}. This system may lead to critical behavior beyond a mean field description\mycite{Gopalakrishnan09,sarang10}, give rise to new glassy phases\mycite{gopalakrishnan2011frustration,strack11, buchhold2013,rotondo2015replica}, and have
potential applications for memory storage\mycite{gopalakrishnan12,torggler17} and optimization problems\mycite{torggler2018quantum}.

The analysis of driven dissipative Dicke model raises many interesting questions. For example, the zero-temperature Dicke model was considered by Emary and Brandes\mycite{Emary2003,Emary2003a} in the framework of classical and quantum chaos. These authors found that the Dicke model (but not the Tavis--Cummings model) has a sharp transition between regular and chaotic motion. Interestingly, in the limit of large $N$, the position of the onset of chaos coincides with the quantum phase transition.
%The effects of drive and dissipation on the chaotic motion of the model evokes many interesting questions.
The relation between quantum chaos and thermalization in the Dicke model was studied for example by Refs.~\cite{lambert2009quantum,altland2012quantum, altland2012, bakemeier2013dynamics,dallatorre2017scale}. To fully access the chaotic regime, it is necessary to go beyond the linear stability analysis reviewed in this report. \jk{In addition to the critical behavior of the open Dicke model discussed in this review, other works have analyzed the behavior of this model from alternate perspectives, such as quantum information approaches\mycite{dey2012information}, large deviation approaches and the $s$-ensemble\mycite{rotondo2018singularities},  and fluctuation-dissipation relations\mycite{mur2018revealing}.}

As we have shown, despite its long history, the Dicke model has continued to reveal new insights about the relation of phase transitions in equilibrium and driven systems.  As a paradigmatic model of many body quantum optics, it continues to play an important role in framing discussions of collective behavior. Given the variety of different directions currently studied experimentally and theoretically, it is likely new understanding will continue to arise from this field in the future.

\let\oldaddcontentsline\addcontentsline% Store \addcontentsline
\renewcommand{\addcontentsline}[3]{}% Make \addcontentsline a no-op

{\bf Acknowledgments} We would like to thank Qing-Hu Chen, {Sebastian Diehl},  Peter Domokos, {Tobias Donner}, Andreas Hemmerich,  Benjamin Lev, Francesco Piazza, Peter Rabl, and Nathan Shammah for reading an earlier version of this manuscript and giving important comments. {P.K. acknowledges support from EPSRC (EP/M010910/1) and the Austrian Academy of Sciences ({\"O}AW). P.K. and J.K. acknowledge support from EPSRC program ``Hybrid Polaritonics'' (EP/M025330/1)}. M.M.R. and E.G.D.T. are supported by the Israel Science Foundation Grant No. 1542/14.

\appendix

\section{Equations of motion and retarded Green's functions%
        \label{sec:aconn}}
{In this appendix we show how to obtain the retarded Green's functions of a set of operators, starting from their Heisenberg equations of motion. %, which can be easily computed in a Master equation approach.
%In this section we explain the full derivation connecting the linear equations of motion to the retarded Green's function of the system in Keldysh notation.
Our approach applies to equations of motion given by the \emph{linear} relation,
\begin{align}
        \label{eq:alin}
        \dot{\vec{v}}(t)=&M \vec{v}(t),
\end{align}
% where $M$ is a constant matrix. This equation is formally solved by
% \begin{align}
%         \label{eq:avt}
%         \vec{v}(t)=&e^{Mt}\vec{v}(0).
% \end{align}
Our goal is to find the corresponding retarded Green's function, defined by
\begin{align}
        \label{eq:argfd}
        G^R_{i,j}(t)=&{-}i\average{\left[v_i\nd(t),v_j\yd(0)\right]}\theta(t).
\end{align}
{We denote the equal-time correlation functions of these operators by a constant matrix $S_{i,j}=\average{\left[v_i\nd(0),v_j\yd(0)\right]}$. In terms
of this matrix, we may write:
\begin{equation}
  \label{eq:1}
  \partial_t G^R_{i,j}(t)
  = - i \delta(t) S_{i,j}
  + M_{i,k}  G^R_{k,j}(t).
\end{equation}
By defining the Fourier transform as $$f(\omega)=\int\limits_{-\infty}^{\infty}\mathrm{d}t\,e^{i\omega t}f(t),$$ we can write Eq.~(\ref{eq:1}) in the matrix form $$({M}+i\omega \mathbbm{1}) {G}^R(\omega) = i {S}.$$ This equations can be explicitly inverted to give
}

% By substituting Eq.~\eqref{eq:avt} in the definition of the retarded Green's function (Eq.~\eqref{eq:argfd}) we find
% \begin{align}
% G^R_{i,j}(t)=&i\average{\left[\sum\limits_k\left(e^{Mt}\right)_{i,k}v_k\nd(0),v_j\yd(0)\right]}\theta(t)\\
% =&i\sum\limits_k\left(e^{Mt}\right)_{i,k}\average{\left[v_k\nd(0),v_j\yd(0)\right]}\theta(t)\\
% =&i\sum\limits_k\left(e^{Mt}\right)_{i,k}S_{k,j}\nd\theta(t), \label{eq:argf}
% \end{align}
% In a vector notation, Eq.~(\ref{eq:argf}) is equivalent to $ G^R(t)=i\theta(t)\left(e^{Mt}S\right)$.
% Having defined the Fourier transformation as
% %with the Fourier transformation defined as
%         $f(\omega)=\int\limits_{-\infty}^{\infty}\mathrm{d}t\,e^{i\omega t}f(t)$,
%         we obtain
% %        we calculate the Fourier transform of Eq.~\eqref{eq:argfv}:
 \begin{align}
         \label{eq:argfo}
         G^R(\omega)=&\left[\omega{\mathbbm{1}}-iM\right]^{-1}S.
 \end{align}
 This expression gives a general connection between the \emph{linear} equations of motions for a set of operators, and the retarded Green's function for the same set of operators.  Note that this expression is valid as long as the equal-time commutators, $S_{i,j}$, are constant in time.}

\bibliographystyle{rsc}
\bibliography{Dicke}

%\newpage
\let\addcontentsline\oldaddcontentsline% Restore

\end{document}